\documentclass[twocolumn,twocolappendix]{aastex631}
\pdfoutput=1
\usepackage{amsmath,amstext}
\usepackage[T1]{fontenc}
\usepackage[figure,figure*]{hypcap}
\usepackage{natbib}
\usepackage{hyperref}
\usepackage{booktabs}
\usepackage[normalem]{ulem}
\usepackage{graphicx}
\usepackage{multirow}
\usepackage[encapsulated]{CJK}
\usepackage[utf8x]{inputenc} 
\usepackage[caption=false]{subfig}
\newcommand{\grizli}{\textsc{gri$z$li}}
\newcommand{\eazy}{\textsc{ea$z$y}}
\newcommand{\pros}{\textsc{prospector}}

\newcommand{\galfit}{\textsc{galfit}}

\newcommand{\source}{A2744-DR2-35602}
\newcommand{\Angstrom}[1]{\AA}

\received{January 9, 2025}
\revised{February 6, 2026}
\accepted{February 26, 2026}
\submitjournal{the Astrophysical Journal}
\shorttitle{The Relic}
\shortauthors{Whitaker et al.}

\begin{document}

\title{Discovery of Ancient Globular Cluster Candidates in the Relic, a Quiescent Galaxy at $z=2.5$}

\correspondingauthor{Katherine E. Whitaker}
\email{kwhitaker@astro.umass.edu}

\author[0000-0001-7160-3632]{Katherine E. Whitaker}
\affiliation{Department of Astronomy, University of Massachusetts, Amherst, MA, USA}
\affiliation{Cosmic Dawn Center (DAWN), Denmark}

\author[0000-0002-7031-2865]{Sam E. Cutler}
\affiliation{Department of Astronomy, University of Massachusetts, Amherst, MA, USA}
\affiliation{Department of Physics \& Astronomy, Tufts University, Medford, MA, USA}

\author[0000-0003-0085-4623]{Rupali Chandar}
\affiliation{Ritter Astrophysical Research Center, University of Toledo, Toledo, OH, USA}

\author[0000-0002-9651-5716]{Richard Pan}
\affiliation{Department of Physics \& Astronomy, Tufts University, Medford, MA, USA}

\author[0000-0003-4075-7393]{David J. Setton}\thanks{Brinson Prize Fellow}
\affiliation{Department of Astrophysical Sciences, Princeton University, Princeton, NJ, USA}

\author[0000-0001-6278-032X]{Lukas J. Furtak}
\affiliation{Department of Physics, Ben-Gurion University of the Negev, Be'er-Sheva, Israel}

\author[0000-0001-5063-8254]{Rachel Bezanson}
\affiliation{Department of Physics \& Astronomy and PITT PACC, University of Pittsburgh, Pittsburgh, PA, USA}

\author[0000-0002-2057-5376]{Ivo Labb\'{e}}
\affiliation{Centre for Astrophysics and Supercomputing, Swinburne University of Technology, Melbourne, VIC, Australia}

\author[0000-0001-6755-1315]{Joel Leja}
\affiliation{Department of Astronomy \& Astrophysics, The Pennsylvania State University, University Park, PA, USA}
\affiliation{Institute for Computational \& Data Sciences, The Pennsylvania State University, University Park, PA, USA}
\affiliation{Institute for Gravitation and the Cosmos, The Pennsylvania State University, University Park, PA, USA}

\author[0000-0002-1714-1905]{Katherine A. Suess}
\affiliation{Department for Astrophysical and Planetary Science, University of Colorado, Boulder, CO, USA}

\author[0000-0001-9269-5046]{Bingjie Wang (\begin{CJK*}{UTF8}{gbsn}王冰洁\end{CJK*})}\thanks{NHFP Hubble Fellow}
\affiliation{Department of Astronomy \& Astrophysics, The Pennsylvania State University, University Park, PA, USA}
\affiliation{Institute for Computational \& Data Sciences, The Pennsylvania State University, University Park, PA, USA}
\affiliation{Institute for Gravitation and the Cosmos, The Pennsylvania State University, University Park, PA, USA}
\affiliation{Department of Astrophysical Sciences, Princeton University, Princeton, NJ, USA}

\author[0000-0003-1614-196X]{John R. Weaver}\thanks{Brinson Prize Fellow}
\affiliation{Department of Astronomy, University of Massachusetts, Amherst, MA, USA}
\affiliation{MIT Kavli Institute for Astrophysics and Space Research, Cambridge, MA, USA}

\author[0000-0002-7570-0824]{Hakim Atek}
\affiliation{Institut d'Astrophysique de Paris, CNRS, Sorbonne Universit\'e, Paris, France}

\author[0000-0003-2680-005X]{Gabriel B. Brammer}
\affiliation{Cosmic Dawn Center (DAWN), Denmark}
\affiliation{Niels Bohr Institute, University of Copenhagen, Copenhagen, Denmark}

\author[0000-0002-1109-1919]{Robert Feldmann}
\affiliation{Department of Astrophysics, Universität Zürich, Zurich, Switzerland}

\author[0000-0003-4264-3381]{Natascha M. F\"orster Schreiber}
\affiliation{Max-Planck-Institut f\"ur extraterrestrische Physik, Garching, Germany}

\author[0000-0002-3254-9044]{Karl Glazebrook}
\affiliation{Centre for Astrophysics and Supercomputing, Swinburne University of Technology, Melbourne, VIC, Australia}

\author[0000-0002-2380-9801]{Anna de Graaff}
\affiliation{Max-Planck-Institut f\"ur Astronomie, Heidelberg, Germany}

\author[0000-0002-5612-3427]{Jenny E. Greene}
\affiliation{Department of Astrophysical Sciences, Princeton University, Princeton, NJ, USA}

\author[0000-0002-3475-7648]{Gourav Khullar}
\affiliation{Department of Astronomy, University of Washington, Seattle, WA, USA}

\author[0000-0001-9002-3502]{Danilo Marchesini}
\affiliation{Department of Physics \& Astronomy, Tufts University, Medford, MA, USA}

\author[0000-0003-0695-4414]{Michael V. Maseda}
\affiliation{Department of Astronomy, University of Wisconsin-Madison, Madison, WI, USA}

\author[0000-0001-8367-6265]{Tim B. Miller}
\affiliation{Center for Interdisciplinary Exploration and Research in Astrophysics (CIERA), Northwestern University, Evanston, IL, USA}

\author[0000-0001-5356-2419]{Houjun Mo}
\affiliation{Department of Astronomy, University of Massachusetts, Amherst, MA, USA}

\author[0000-0002-8530-9765]{Lamiya A. Mowla}
\affiliation{Department of Physics and Astronomy, Wellesley College, Wellesley, MA, USA}

\author[0000-0003-2804-0648]{Themiya Nanayakkara}
\affiliation{Centre for Astrophysics and Supercomputing, Swinburne University of Technology, Melbourne, VIC, Australia}

\author[0000-0002-7524-374X]{Erica J. Nelson}
\affiliation{Department for Astrophysical and Planetary Science, University of Colorado, Boulder, CO, USA}

\author[0000-0002-0108-4176]{Sedona H. Price}
\affiliation{Space Telescope Science Institute, Baltimore, MD, USA}
\affiliation{Department of Physics \& Astronomy and PITT PACC, University of Pittsburgh, Pittsburgh, PA, USA}

\author[0000-0001-9705-2461]{Francesca Rizzo}
\affiliation{Kapteyn Astronomical Institute, University of Groningen, Groningen, The Netherlands}

\author[0000-0002-8282-9888]{Pieter van Dokkum}
\affiliation{Department of Astronomy, Yale University, New Haven, CT, USA}

\author[0000-0003-2919-7495]{Christina C.\ Williams}
\affiliation{NSF National Optical-Infrared Astronomy Research Laboratory, Tucson, AZ, USA}
\affiliation{Steward Observatory, University of Arizona, Tucson, AZ, USA}

\author[0000-0002-5564-0254]{Yanzhe Zhang}
\affiliation{Department of Astronomy, University of Massachusetts, Amherst, MA, USA}

\author[0000-0001-6454-1699]{Yunchong Zhang}
\affiliation{Department of Physics \& Astronomy and PITT PACC, University of Pittsburgh, Pittsburgh, PA, USA}

\author[0000-0002-0350-4488]{Adi Zitrin}
\affiliation{Department of Physics, Ben-Gurion University of the Negev, Be'er-Sheva, Israel}

\begin{abstract}\noindent
Globular clusters (GCs) are some of the oldest bound structures in the Universe, holding clues to the earliest epochs of star formation and galaxy assembly.  However, accurate age measurements of ancient clusters are challenging due to the age-metallicity degeneracy. Here, we report the discovery of 36 compact stellar systems within the 'Relic', a massive, quiescent galaxy at $z=2.53$. The Relic resides in an overdensity behind the Abell~2744 cluster, with a prominent tidal tail extending towards two low-mass companions. Using deep data from the UNCOVER/MegaScience JWST Surveys, we find that clusters formed in age intervals ranging from 8~Myr up to $\sim2$~Gyr, suggesting a rich formation history starting at $z\sim10$. While the cluster-based star formation history is broadly consistent with the high past star formation rates derived from the diffuse host galaxy light, one potential discrepancy is a tentative $\sim2-3\times$ higher rate in the cluster population for the past Gyr.  Taken together with the spatial distribution and low inferred metallicities of these young-to-intermediate age clusters, we may be seeing direct evidence for the accretion of star clusters in addition to their early in situ formation. The cluster masses are high, $\sim10^6-10^7~M_{\odot}$, which may explain why we are able to detect them around this likely post-merger galaxy. Overall, the Relic clusters are consistent with being precursors of the most-massive present-day GCs. This unique laboratory enables the first connection between long-lived, high-redshift clusters and local stellar populations, offering insights into the early stages of GC evolution and the broader processes of galaxy assembly.
\end{abstract}


\section{Introduction}

A hallmark of the standard $\Lambda$CDM cosmological framework asserts that galaxies assemble piecewise, continually fueling new generations of star formation via in-flowing gas and the accretion of smaller satellite galaxies. Understanding this process is key to understanding how modern massive galaxies, including our own Milky Way, form and evolve. Direct observations of compact star clusters
at high redshift, when the Universe is only a few billion years old, therefore provide critical evidence necessary to map out the hierarchical growth process at the earliest times.
This is because ultimately most star formation occurs in clusters \citep{Lada2003, Krumholz2019}, with a significant fraction therein occurring in the compact, star-forming clusters observed at early epochs \citep[e.g.,][]{Adamo2020}. Therefore, the problem of understanding the structural evolution of galaxies in a hierarchical framework is inherently linked to the density and resulting lifetimes of their building blocks, stellar associations.

Little is known about how and when long-lived, dense globular clusters (GCs) in the Milky Way and nearby galaxies formed. The theoretical consensus is that GCs generally emerge from regions within the interstellar medium with violent conditions, including high gas densities, turbulent velocities, and hence extreme gas pressures  \citep[see e.g.,][]{Kruijssen2014}. GCs are thought to form primarily in situ, while the stripping of accreted satellites provides an important channel for assembling GC systems in massive galaxies \citep{BrodieStrader2006, Pota2013, LiGnedin2014, Harris2015, ForbesRemus2018, Dolfi2021, Kluge2023}. Observationally, both massive elliptical galaxies and spiral galaxies with prominent bulges have long been known to host two populations of GCs, one that is metal-rich and red, and another that is metal-poor and relatively bluer in color \citep[e.g.][]{Peng2004, Faifer2011, Brodie2012, Usher2012, Pota2013}. 

The origin of the bimodality in GC colors and metallicities remains the most outstanding question of GC formation.  
The metal-rich GCs have been suggested to have formed either in galaxy mergers \citep{Ashman1992, Beasley2002, Newton2024, DeLucia2024} or coevally with the bulk of the stellar populations within the host galaxy \citep{Forbes1997, Strader2005, Chen2024, DeLucia2024}.  
Those metal-rich clusters that reside in the halo in particular may have initially formed in protodisks but are thrown out by merger-driven violent relaxation \citep{Hopkins2009}.  
On the other hand, the metal-poor GCs may have formed during the collapse of protogalactic clouds \citep{Forbes1997, Beasley2002}, which is suggested to have been truncated by reionization \citep{Strader2005}.
Another explanation for the bimodality of the GC population is that it is a natural result of hierarchical galaxy formation, in which the metal-poor GCs originate in cannibalized dwarf galaxies \citep[e.g.,][]{Cote1998, Hilker1999, KisslerPatig2000, Mackey2004}.

Although the common interpretation is that red, metal-rich GCs predominantly form during the in situ phase and blue, metal-poor GCs during an accretion phase of massive galaxy formation \citep[e.g.,][]{Peng2004, Pota2013, Kluge2023}, there is not a strict one-to-one connection \citep{ForbesRemus2018}. That said, there may be no need for separate formation processes at all, as cluster disruption may explain the bimodal GC metallicity distribution \citep{Pfeffer2023}.  While debate continues on the where and the how of GC formation, the age distribution of GCs in galaxies at $z>1$ can directly address these questions given that massive star clusters trace major star-forming episodes in their parent galaxies.      

With the first Webb Deep field, SMACS0723, the James Webb Space Telescope \citep[JWST;][]{Gardner2023} imaging and slitless spectroscopy 
immediately revealed GC candidates in a strongly lensed, $z=1.378$  galaxy \citep[``The Sparkler'',][]{Mowla2022, Claeyssens2023}. Spectrophotometric fitting suggests that these compact, red sources are likely ancient GCs that formed between $7<z<11$. For context, GCs in the Milky Way have ages ranging from $\sim$11-13 Gyrs, implying that the oldest Milky Way GCs formed at $z\sim8-10$ \citep{Beaulieu2001, Momany2003, DeAngeli2005, MarinFranch2009, Leaman2013, delaFuente2015}.  
However, further analysis of the GC candidates in the Sparkler confirm that while most were likely formed very early in the Universe, there is a significant spread in metallicity that calls into question their assembly history \citep{Adamo2023}.  

Robustly constraining the ages of GCs in the Milky Way, nearby galaxies, and even the Sparkler remains challenging because these systems are uniformly old. Beyond a few gigayears, the evolution of integrated colors and spectral features with age becomes increasingly subtle, leading to strong degeneracies with metallicity, abundance patterns, and stellar evolution assumptions \citep{Conroy2010}. For integrated-light and spectrophotometric analyses, this typically results in absolute age uncertainties of order $\sim$1–3 Gyr for well-studied Galactic GCs
\citep[e.g.,][]{DeAngeli2005,MarinFranch2009}. At higher redshift, however, the same absolute age differences correspond to a much larger fraction of the cosmic age, such that dense multiband photometry can provide meaningful constraints on formation timescales \citep[e.g., as confirmed by][]{Belli2019}, even when metallicity remains poorly constrained. High signal-to-noise ratio (SNR) spectroscopy can further improve age constraints, but at $z \gtrsim 1$ degeneracies with metallicity and abundance patterns often limit progress \citep[e.g.,][]{Kriek2009, Kriek2016,Beverage2025}, whereas at lower redshift ($z\lesssim1$) uncertainties associated with horizontal branch morphology and late stellar evolutionary phases become dominant \citep{Conroy2010}. As a result, it is difficult to determine whether GC ages truly approach the age of the Universe, even at intermediate redshifts such as $z\sim1.5$ ($t_{\mathrm{univ}}$=4.5 Gyr). 
 
With the first few drips of data from JWST, the Sparkler rapidly became the best known example in the literature to date of a galaxy containing GC candidates at $z>1$ -- the infrared sensitivity of this facility immediately opened a brand new discovery space for studying GC formation at early times. At higher redshifts, however, age constraints become intrinsically more informative, as the same absolute age differences correspond to a larger fraction of the cosmic age. Recent studies of early quenched galaxies have therefore achieved substantially improved constraints on formation timescales \citep[e.g.,][]{Carnall2023, Glazebrook2024, deGraaff2024}.

There is a growing literature of JWST studies of the stellar populations of high redshift clumps, where age-dating is more straight forward owing simply to the young age of the Universe itself \citep{Claeyssens2023, Messa2024a, Messa2024b, Fujimoto2024, Welch2023, Vanzella2022b, Vanzella2023, Bergamini2023}. Possible GC candidates may even now be identified as early as $z=8-10$ \citep{Adamo2024, Bradley2024, Mowla2024}.   
Note that some published works on cluster systems at high redshift use the term 'clump' to refer to a wide range of stellar systems, extending from the most compact sources that could still have sizes of a few 100~pc up to much larger ($\approx$kpc-scale) super-complexes that are spatially resolved \citep[e.g.,][]{ForsterSchreiber2011, Guo2015, Guo2018}. 
However, measurements in highly lensed systems can provide improved constraints for the sizes of stellar sources, reaching half-light radii of just a few pc \citep[and even measured sizes in a handful of cases;][]{Vanzella2022a, Vanzella2023}; we therefore will refer to the compact (unresolved) stellar sources which have estimated ages of at least 10~Myr as `clusters' or `globular clusters', since they have likely survived early dispersal and can be considered gravitationally bound.

Hunting for the oldest bona fide GC candidates that have well passed this threshold to survive disruption limits the redshift range effectively to $z<5$, whereas complexities in constraining age bookend this limit to $z\gtrsim2$ (i.e., $1<t_{\mathrm{univ}}<3$ Gyr). 
Studies in this redshift space have discovered many star clusters, but all are limited to ages $<300$ Myr \citep{Vanzella2022a, Messa2024a}. Moreover, the host galaxies all appear to be actively star-forming, similar to the rich history of earlier Hubble Space Telescope giant clump studies \citep[e.g.,][]{ForsterSchreiber2011, Wuyts2012, Guo2012, Guo2015, Guo2018}. A recent comprehensive analysis of star clusters within galaxies behind the Abell~2744 cluster expands this parameter space \citep{Claeyssens2024}, though only 3 out of 1956 clusters fall in the relevant age ($>800$ Myr) and redshift range ($z>2.5$) that would correspond to first GC candidates unambiguously capable of surviving to the present day. 

In this work, we report the discovery of a remarkable laboratory: a quenched, $\log(M_\star/M_\odot)=10.86^{+0.04}_{-0.16}$ galaxy{\footnote{This value is corrected for magnification.}} (\source{}) at $z_{\rm phot}=2.53^{+0.12}_{-0.36}$ with an extended smooth light profile, scattered with compact stellar systems that may be the earliest known GCs (Figure \ref{fig:image}).
\source{}, hereafter ``The Relic,'' is a classical quiescent galaxy first identified in the Cycle 1 JWST UNCOVER Treasury Program \citep[JWST-GO-2561; PIs: Labb\'{e}/Bezanson,][]{Bezanson2024} photometric catalog \citep{Weaver2024}. The Relic is a remarkably bright (F444W = 20.52 ABmag) quiescent target ($\log({\rm sSFR/yr^{-1}})=-9.61^{+0.51}_{-0.15}$, i.e., $\sim1$ dex
below the main sequence from \citet{Speagle2014, Leja2022}. 

The Relic features a large, extended de Vaucouleurs-like light profile, with a tidal tail connecting it to two lower-mass quiescent galaxies consistent with the same redshift.  The system of galaxies reside in the northwest outskirts of the strong lensing cluster Abell 2744, outside of the caustic.  As such, the Relic and its two neighbors have modest magnifications of $\mu\sim2.7$ \citep{Furtak2023}, with ultra-deep imaging in the complete suite of medium-band and broadband JWST/NIRCam filters from 0.7-4.4$\mu$m when also including Cycle 2 data from JWST-GO-4111 \citep[PI: Suess;][]{Suess2024}.  While ultra-deep follow-up IFU spectroscopy of the Relic was acquired on October 28, 2024 and December 15-16, 2024 through a Cycle 3 program (JWST-GO-6405, PIs: Cutler/Whitaker), we focus herein on the photometric properties of this target over a significantly wider field of view relative to the two 3$^{\prime\prime}$ IFU footprints.  The purpose of this paper is to report the physical properties of the GC candidates derived from the JWST imaging dataset, whereas spectroscopic redshift confirmation and a more detailed resolved spectral analysis will come with the Cycle 3 data.  

The paper is organized as follows: Section~\ref{sec:data} presents the data reduction, details on the source detection, photometry, and photometric redshift and spectral synthesis modeling is found in Section~\ref{sec:methods}, the resulting stellar population trends are summarized in Section~\ref{sec:results}, and we place our findings in the context of our current understanding of GC formation in Section~\ref{sec:discussion}.  We assume a \cite{Chabrier2003} initial mass function (IMF) and a standard concordance cosmology: $H_0=70~{\rm km~s^{-1}~Mpc^{-1}}$, $\Omega_M=0.3$, and $\Omega_\Lambda=0.7$. All magnitudes in this paper are expressed in the AB system \citep{Oke1974}, for which a flux $f_\nu$ in $10\,\times\,$nJy ($10^{-28}$~erg~cm$^{-2}\,$s$^{-1}\,$Hz$^{-1}$) corresponds to AB$_\nu=28.9-2.5\,\log_{10}(f_\nu/\mu{\rm Jy})$.

\begin{figure*}
    \centering
    \includegraphics[width=\linewidth]{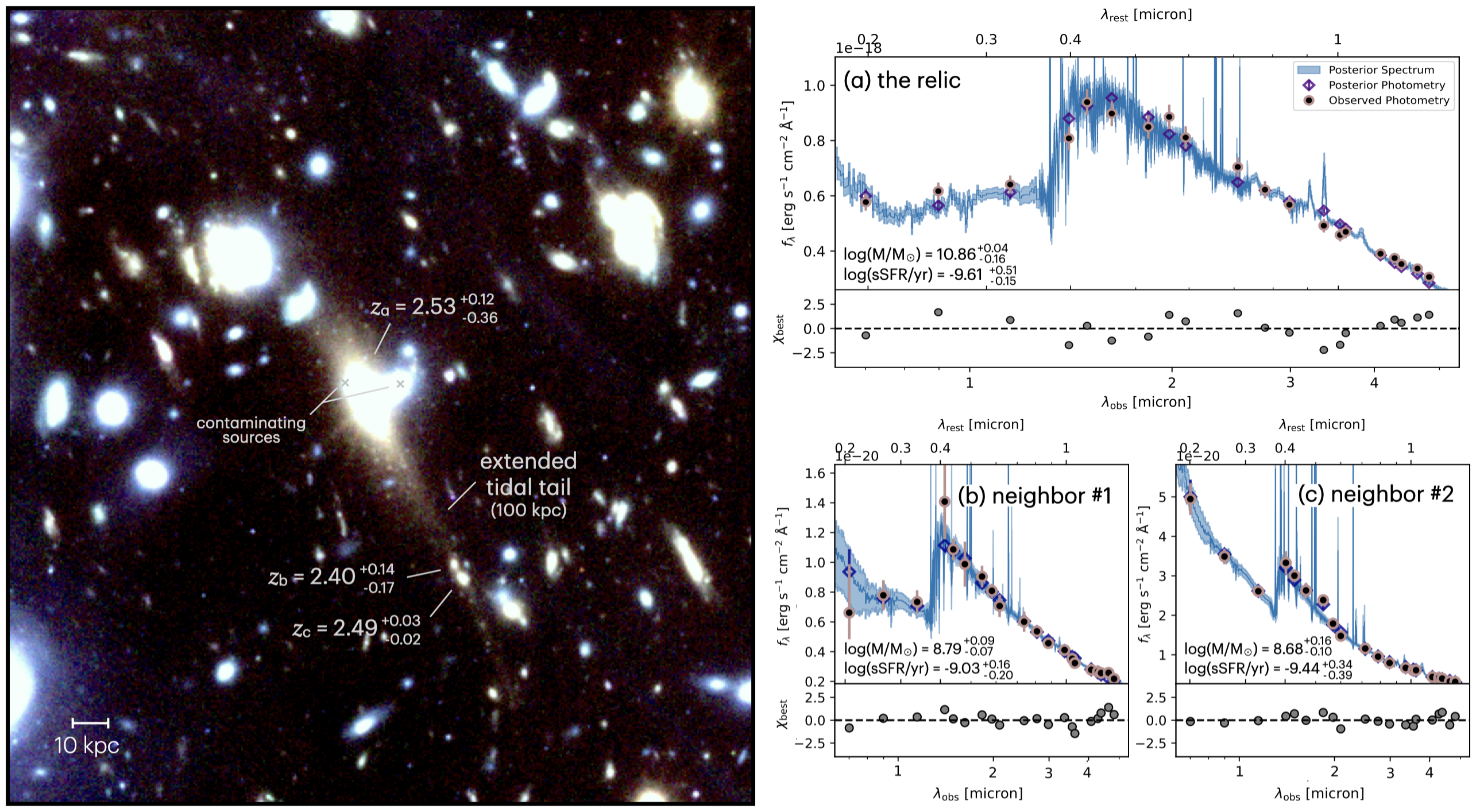}
    \caption{\textbf{Left:} Composite color image (F277W, F356W, and F444W) of the Relic (a) and two lower mass neighboring galaxies to the bottom right (b and c) connected by a diffuse tidal tail that extends roughly 100 kpc (40 kpc in the source plane). \textbf{Right:} The spectral energy distributions of the three galaxies are all consistent with being quiescent.  The photometric redshifts for all three are consistent within the uncertainties.  The scale bar is in the observed plane, whereas the source plane scale is roughly the same perpendicular to the main axis of the Relic and a factor of 2.5 smaller parallel to the main axis. }
    \label{fig:neighbors}
\end{figure*}

\section{Data}\label{sec:data}

The Relic ($\alpha$,$\delta$ = 3.5277082,-30.3665864) was first identified in the extended coverage afforded by the UNCOVER treasury program \citep[JWST-GO-2561; PIs: Labb\'{e}/Bezanson][]{Bezanson2024}, including 45 square arcminutes of near-infrared imaging of the strong gravitational lensing cluster Abell 2744. The original UNCOVER NIRCam image set includes F090W, F115W, F150W, F200W, F277W, F356W, F410M, and F444W, later augmented by 12 additional bands (F070W, F140M, F162M, F182M, F210M, F250M, F300M, F335M, F360M, F430M, F460M, and F480M) through the Cycle 2 MegaScience Program \citep[JWST-GO-4111, PI: Suess;][]{Suess2024} and additional integration in F070W and F090W filters through the Cycle 2 ALT Program \citep[JWST-GO-3516, PIs: Matthee, Naidu;][]{Naidu2024}.  The combined data set is one of the deepest-to-date (when augmented by strong lensing) publicly available Webb surveys \citep{Bezanson2024,Suess2024}. 

The data presented in this paper were originally obtained from the Mikulski Archive for Space Telescopes (MAST) at the Space Telescope Science Institute.  All images correspond to the v7\footnote{\url{https://dawn-cph.github.io/dja/imaging/v7/}} mosaics reduced by \grizli{} \citep{grizli} and rescaled to a 40 mas pixel scale in all available filters. The observations can all be accessed at \url{https://doi.org/10.17909/nftp-e621}, with the original photometric catalogs available at \url{https://doi.org/10.5281/zenodo.8199802}. While the UNCOVER mosaics formally have bright cluster galaxies, intracluster light, and sky background removed, as described in \citet{Weaver2024}, the Relic is sufficiently far from the cluster core that these effects are not important.  We do, however, note two moderately bright sources within an arcsecond of the main target and describe how we remove their contaminating light in Section~\ref{sec:methods}.

We adopt an updated version of the strong lens model from \cite{Furtak2023}. The Relic is located in a moderate-magnification region with an average magnification factor of $\mu=2.75^{+0.12}_{-0.28}$ (with $\mu_{\mathrm{radial}}=1.10^{+0.01}_{-0.02}$ and  $\mu_{\mathrm{tangential}}=2.49^{+0.09}_{-0.12}$). All photometric measurements have been corrected for this flux boost, assuming the average magnification value.
While the statistical uncertainty remains small in this low-magnification regime, systematic modeling uncertainties are expected to be on the order of 20\% \citep{zitrin15,atek24}. These uncertainties have been incorporated into the photometry.

We adopt a redshift of $z=2.53$ herein for the Relic and all associated cluster modeling.  With the 20-band photometry, the redshift probability distribution for the Relic returns three viable solutions: (1) $z=2.28$, (2) $z=2.53$ (our fiducial value), and (3) $z=2.68$.  The $z=2.53$ solution is favored when fitting the photometry using Prospector-$\beta$ \citep{Wang2023, Wang2023b}, whereas EaZY favors the $z=2.28$ solution. Medium-resolution ($R\sim1600$) grism spectroscopy in the range of $\sim$3-4$\mu$m is publicly available for the Relic \citep[JWST-GO-3516; PIs: Matthee, Naidu;][]{Naidu2024}; however, only one roll angle is usable, and the spectrum extracted for the Relic is noisy.  From this spectrum, we confirm that no emission lines are present, supporting the quiescent nature of the Relic.  We also find tentative evidence for a broad Pa$\gamma$ ($\lambda_{\mathrm{rest}}$=1.094$\mu$m) absorption feature consistent with a spectroscopic redshift of $z=2.524$.  However, we note that if Pa$\gamma$ were observable, we also should have seen Pa$\delta$, which was not detected.  While a proper reduction of the recent deep IFU data is not completed, we see tentative evidence for a break consistent with $z=2.53$.  Finally, in contrast, there does appear to be one cluster (ID8) that may have emission-line boosting that is only possible with the $z=2.68$ solution; while this cluster may not be physically associated with the Relic, the solution of $z=2.53$ is within the posterior distribution function.    

Other evidence supporting the $z=2.5$ solution includes the discovery of a large overdensity of $\sim$100 quiescent galaxies at $z=2.5$ within the Abell~2744 field of view \citep[][]{Naidu2024,Pan2025}.  There exists a faint extended tidal tail to the south of the Relic, connecting the main target with two lower-mass (log(M/M$_{\odot}$)$\sim$9) quiescent galaxies about 70 kpc away ($\sim$30 kpc in the source plane). Figure~\ref{fig:neighbors} shows the spectral energy distributions and best fit models for these sources adopting the Prospector-$\beta$ physical model with a nonparametric star formation history \citep[SFH;][]{Wang2023b}.  Their photometric redshifts, $z_{\rm phot,b}=2.40^{+0.14}_{-0.17}$ and $z_{\rm phot, c}=2.49^{+0.03}_{-0.02}$, are consistent with the Relic having recently passed by, with a tidal stream marking the past trajectory. All together, while we find ample evidence to support the $z=2.53$ solution as our fiducial value (and a spectroscopic redshift will be coming soon), we are careful to test the impact of alternative redshift solutions on our conclusions for all subsequent analyses. 

\section{Methods}\label{sec:methods}

\begin{figure*}
    \centering
\includegraphics[width=\linewidth]{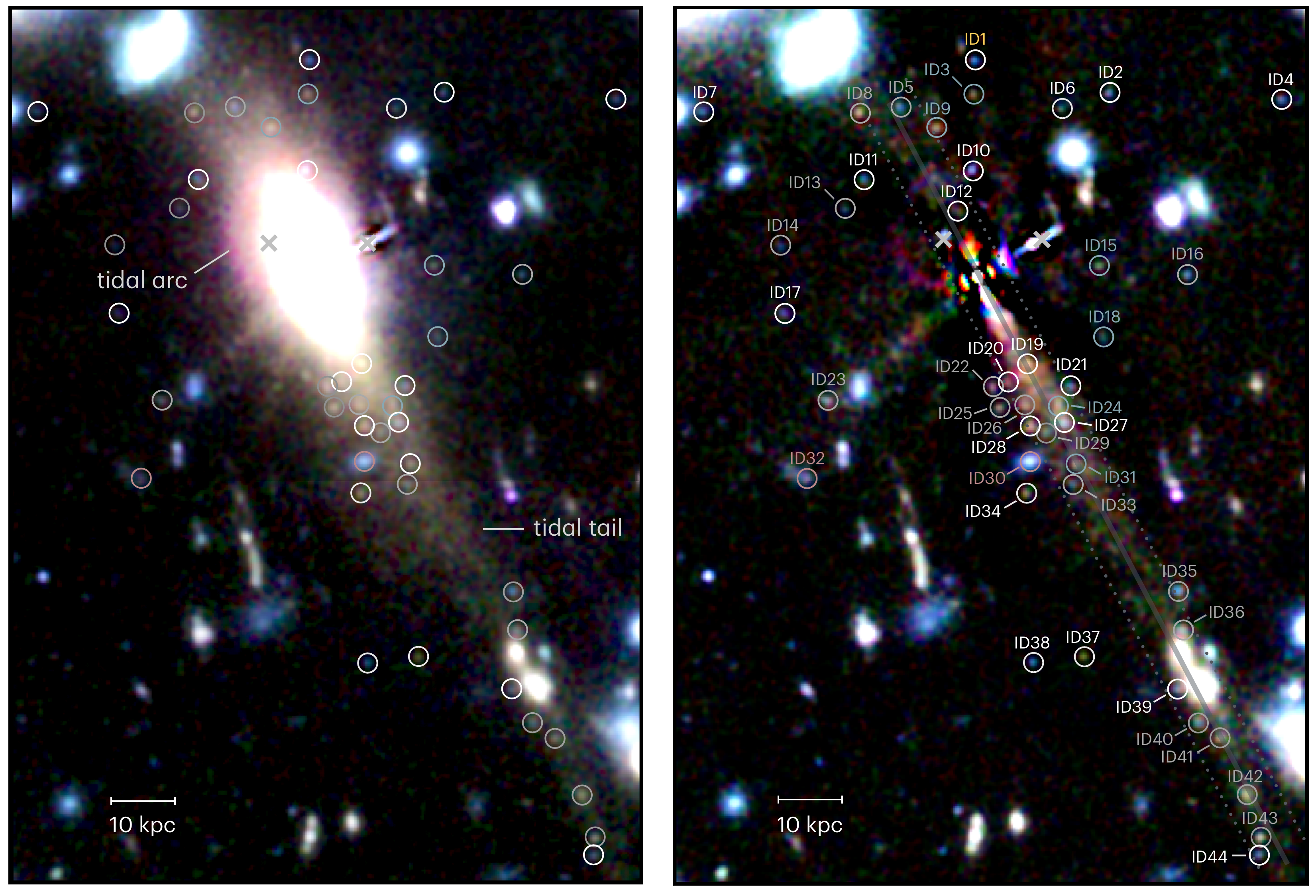}
\caption{Composite color image (F277W, F356W, and F444W) of the Relic before (left) and after (right) subtracting the main host galaxy. All point sources with a photometric redshift consistent with the main target ($z_{\mathrm{phot}}=2.53$) and SNR$>$3 in F444W are encircled, shown as white if the best fit simple stellar population has a reduced $\chi^{2}<2$ and grey where reduced $\chi^{2}>2$. ID30 and ID32 (mauve) are removed from the final sample due tentative evidence suggesting extended sizes (see Section~\ref{sec:sizes}). ID3, ID9, ID15, ID18, ID24, and ID31 (teal) only meet the SNR requires with an alternative photometry approach and are therefore not included in the final sample. The scale bar is in the observed plane, whereas the source plane scale is roughly the same perpendicular to the main axis of the Relic and a factor of 2.5 smaller parallel to the main axis.}
    \label{fig:image}
\end{figure*}

\subsection{Cluster Detection}
\label{sec:detection}

Figure~\ref{fig:image} shows the composite color image (F277W, F356W, and F444W) of the Relic before (left) and after (right) subtracting the smooth elliptical light profile of the host galaxy. 
Sources are detected by running Source Extractor \citep{Bertin1996} on a detection image, which is created by subtracting the original images (Figure~\ref{fig:image}, left) by a median filtered image version in 4 filters (F200W, F277W, F356W, F444W) and  combining them together \citep[e.g.,][]{Goudfrooij2001}. The smoothing scale is set to 9 pixels, but we note that a range from 7 to 10 pixels is acceptable and identifies the same sources unambiguously.  In order to robustly identify clusters in the vicinity of the Relic, we experiment with first removing two moderately bright nearby galaxies (marked with crosses in Figure~\ref{fig:image}) before median subtracting the smoothed image versus the outcome when not subtracting these neighbors. While the removal of these nearby sources following the methodology detailed in \citet{Weaver2024} (i.e., subtracting smooth elliptical light profiles) leaves residuals at the galaxy core for the brighter of the two, this is a trade off with the benefit of also removing extended light that could contaminate the cluster photometry.  We also experiment with running the source detection directly on the galaxy-subtracted image (Figure~\ref{fig:image}, right), but find that we can more robustly identify compact sources when using the median smoothing technique described above. 
We detect 38 sources above a signal-to-noise ratio (SNR) threshold of 3 in F444W and 2 in F200W, with an additional 6 tentative sources identified when adopting alternative background subtraction techniques (Section~\ref{sec:photometry}).  We proceed to vet this sample through an analysis of their spectral energy distributions (SED), as described in the following sections. 

Regardless of the detection approach adopted, no new candidate clusters are discovered close to the bright center of the Relic (see Figure~\ref{fig:image}).  As we consider the region within the central $\lesssim$10 kpc radius ($\sim$4 kpc in the source plane along tidal tail axis) to be contaminated by residual features with minimal contrast possible between the host light and candidate clusters, we can neither confirm nor rule out the absence of clusters within this galactocentric radius.  

\subsection{Cluster Photometry}
\label{sec:photometry}

Before extracting photometry of the detected point sources, all images are homogenized to a common resolution matching the F444W point spread function (PSF).  We use the same convolution kernels as developed and extensively tested in \citet{Weaver2024} and \citet{Suess2024}.  The kernels are derived from matching empirical PSFs built from stars selected across the UNCOVER footprint using \texttt{pypher} \citep{Boucaud2016}.  Due to the dither pattern and conditions at the time of observations, the empirical PSFs are broader than those produced by WebbPSF.  Thus, the use of an empirical PSF is particularly important when deriving the morphologies and/or colors of compact objects.  

We perform photometry on the PSF-convolved images within circular apertures of 0.16$^{\prime\prime}$ diameter (4 pixels) owing to the compact nature of the clusters (see Section~\ref{sec:sizes} for more details); such a small aperture maximizes the SNR recovered for the extracted photometry of compact sources. Given the complex background light surrounding the GC candidates, we opt to use a local background subtraction within an annulus of 0.24-0.28$^{\prime\prime}$ diameter (6-7 pixels) for the photometric analysis, after both neighboring galaxies have been subtracted out. When subtracting a local background, we first determine the background contribution per pixel by multiplying the flux enclosed within the annulus by the ratio of the area of the central 4 pixel aperture relative to that of the annulus. This value is subtracted from the aperture flux. In order to correct to total flux, we estimate the aperture correction by measuring the amount of light from F444W UNCOVER empirical PSF curve of growth that falls outside of a radius of 0.08$^{\prime\prime}$. However, we require an adjustment to the aperture correction when correcting to total flux to account for the PSF wings subtracted within the background annulus. Using our F444W curve of growth, we determine the fraction of the light enclosed in the annulus and scale this by the ratio of the aperture-to-annulus area.  This value is then subtracted from the original fraction of enclosed light within the aperture, increasing the correction factor by 30\% to a value of 3.48 for all PSF homogenized filters.  All fluxes are corrected for galactic extinction, with a maximum correction of 0.02 ABmag in F070W.  
Finally, the flux densities are further systematically down-weighted by the magnification factor at the location of the Relic, $\mu=2.75$. 

We experiment with two alternative approaches to the cluster photometry.  First, we perform photometry on a globally background-subtracted image with the host galaxy light of the Relic removed using the elliptical modeling approach for bCGs described in \citet{Weaver2024}, as shown in the right panel of Figure~\ref{fig:image}). While a local background subtraction may more robustly handle any residual features that the host galaxy model does not capture, potentially leading to biased flux estimates, this default approach runs the risk of over-subtracting the background if the host galaxy light has a steep gradient.  Modeling out the host galaxy light adds six additional sources to the sample owing to the slightly higher flux values and consequently SNRs (teal circles in Figure~\ref{fig:image}; ID3, ID9, ID15, ID18, ID24, and ID31); while we note these sources, we do not include them in our final sample. 

Second, we test a hybrid approach where we add a local background subtraction to the host galaxy-subtracted analysis. As the results are similar to the default local background estimates, this tells us that the host galaxy light gradient is not the dominant effect here. Instead, the host subtraction residuals have comparable structure on these scales, so the local background annulus is responding similarly in both cases.  We present results from this alternative in Section~\ref{sec:modeling}, noting that the main conclusions of this work remain unchanged.

While performing photometric validation checks, we noticed that the short-wavelength (SW) medium-band images had mild residual striping effects in the background.  While the variations are mild, such background variations become important when measuring photometry for objects close to the noise level, and even more so for intrinsically red clusters that are fainter in these blue SW filters. We therefore increase the error bars for all SW medium-band flux density measurements (F140M, F162M, F182M, F210M) by 30\% and set the flux to a null value if the SNR is less than unity (i.e., flux = NaN).  We experimented with different assumptions (i.e., boosting the flux uncertainties higher, removing all medium-band filters, etc), but this did not significantly change the results. 

The aperture photometry of the Relic itself was performed following the methodology outlined in \citet{Weaver2024} and adopted from the publicly released ``super'' photometric catalog.  Briefly, aperture photometry is performed on the bCG-subtracted images that are all homogenized to the F444W PSF. In the ``super'' catalog, the aperture chosen scales with the spatial extent of the object, ranging from 0.32 arcseconds upwards to 1.4 arcseconds in diameter.  The aperture diameter adopted for the Relic is 0.48 arcseconds.  The flux is then scaled to total through aperture corrections that first scale up all filters by the same amount based on the ratio of the flux within the larger Kron radius relative to the circular aperture, and a second correction to account for missing light outside the Kron radius inferred from the empirical F444W PSF curve of growth analysis.  In this approach, the spectral shape and colors of the Relic correspond to the light within the 0.48 arcsecond diameter.  The main host galaxy photometry therefore does not overlap spatially with any clusters detected, as they are all located beyond an arcsecond from the center, and can be treated as a distinct photometric analysis.  

\subsection{Constraints on Cluster Sizes}
\label{sec:sizes}

\begin{figure}
    \centering
\includegraphics[width=0.97\linewidth]{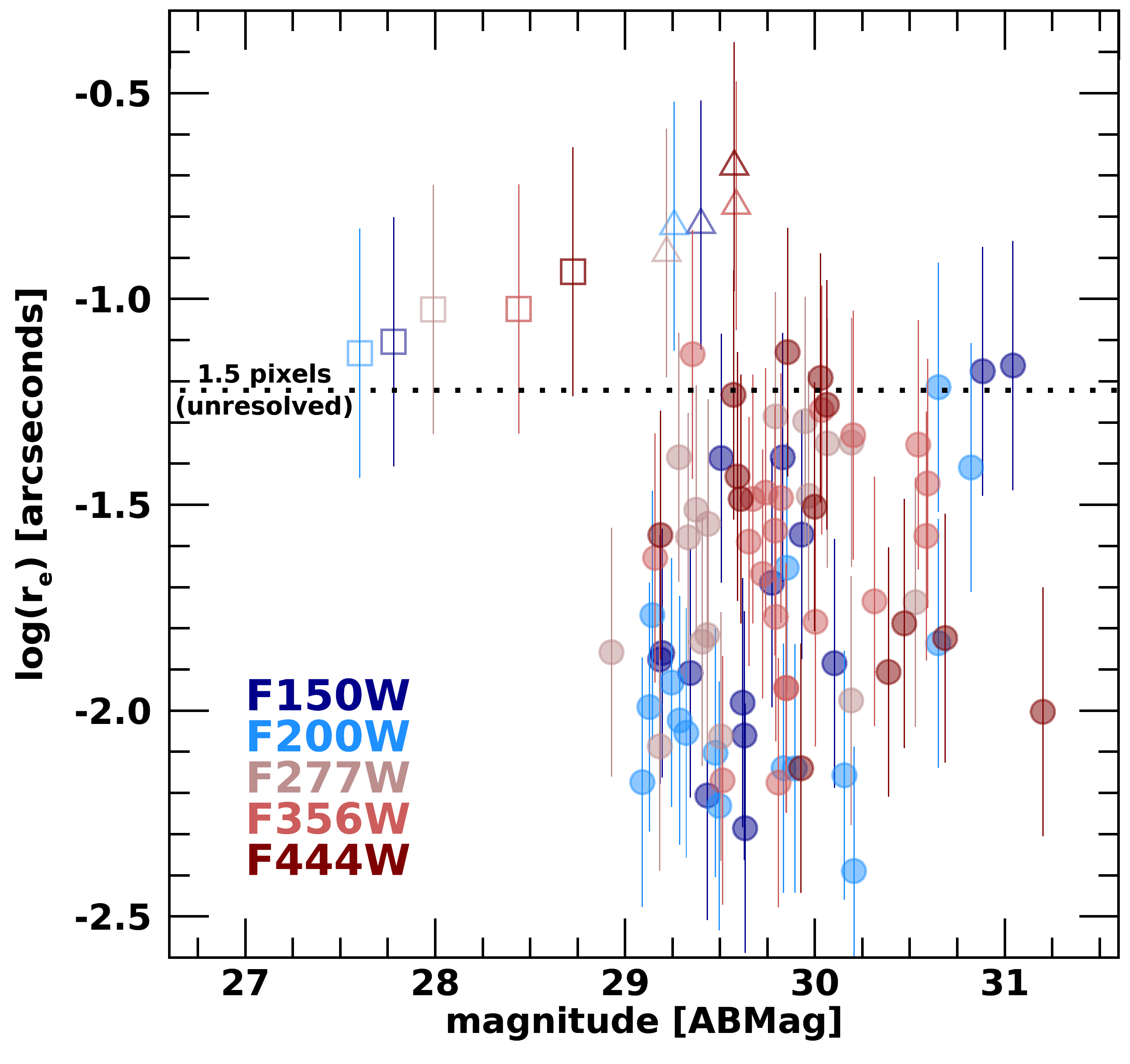}
\caption{Measurements of the effective radius for cluster candidates in the five broadband filters redward of the Balmer break at $z=2.5$ reveal that all sources with robust fits (roughly 50\% of the sample) are consistent with unresolved within the 16th to 84th percentiles of the posterior distribution. Given the faint magnitudes of the candidates, we assume a limit of 1.5 pixels (0.06$^{\prime\prime}$) to define this threshold. Measurements are color-coded by the filter they represent. ID30 (square symbols) is removed from subsequent analyses due to the combined brightness and tentatively extended size.  ID32 (triangle symbols) is removed due to the extended size.}
    \label{fig:sizes}
\end{figure}

In order to demonstrate that the cluster candidates are unresolved, we model the light for all sources in the  five deep broadband filters redward of the Balmer break (F150W, F200W, F277W, F356W, and F444W). We perform a single Sersic profile fit on the original image (before PSF homogenization) at a pixel scale of 0.04$^{\prime\prime}$/pix using \texttt{pysersic} \citep{Pasha2023}. We apply broad, physically motivated priors on structural parameters to allow flexible modeling while preventing unphysical solutions, including a prior range of ([0,10] pix) for effective radius and the default auto prior in \texttt{pysersic} for S\'ersic index and axis ratio, with the position angle unconstrained. While we attempt to model all clusters, we find that the fits more often fail when sources are embedded within the diffuse tidal tail light as the flat sky background term adopted in \texttt{pysersic} cannot robustly handle any gradients in the residuals.  Luckily, such a failure mode returns spectacularly large, extended best fit models that are easy to identify: bad fits are flagged when either r$_e$ is outside of ([0.01, 10] pix) and/or n is outside of ([0.2, 8]), if the model centroid is $>$10 pixels away from that of the catalog, or if the model magnitude deviates by $\geq$3 ABmag from that in the catalog. We recover reliable fits in at least one filter for 22 clusters total.  While this morphological analysis therefore only includes $\sim$60\% of the sample, it is likely the best we can do given the faintness of these clusters candidates and the background contamination.  

For the cluster candidates successfully modeled -- specifically, including 16 clusters (F150W), 18 (F200W), 19 (F277W), 22 (F356W), and 15 (F444W), respectively -- we find a median effective radius and normalized median absolute deviation that ranges from 0.04$\pm$0.04 arcseconds in F150W to 0.12$\pm$0.13 arcseconds in F444W (not correcting for magnification). 
Figure~\ref{fig:sizes} presents the inferred sizes in the five broadband filters. We define a threshold of 1.5 pixels (0.06 arcseconds) as the limit below which a source can be considered unresolved, requiring that $>$50\% of the posterior be below this threshold in at least one filter. For the few cases where the median lies just above our threshold, the source must have a measurement safely below in another filter.  This threshold is  motivated by extrapolating results to fainter magnitudes from the mock recovery tests presented in \citet{Cloonan2026} and related analyses performed on stars versus `Little Red Dots' \citep{Hviding2025,DeGraaff2025}.

We find that 36 out of 38 sources (circles in Figure~\ref{fig:sizes}) are consistent with being unresolved given our criteria.  There exist two moderately bright, blue sources which show tentative evidence for extended light profiles. ID30 (squares) is located moderately close to the tidal tail with an effective radius just above our threshold; while the inferred physical properties are not unphysical (i.e., log(M$_{\star}$/M$_{\odot}$)$\sim$6.95$^{+0.08}_{-0.04}$ and an age of 8.3$^{+0.6}_{-0.3}$ Myr; see methods in Section~\ref{sec:modeling}), we conservatively exclude this source from subsequent analyses. The other outlier, ID32 (triangles), resides in the halo with a visible blue ``smudge'' attached to it.  While the fit is formally resolved, we caution that this may instead be two unresolved clusters.  To remain conservative, we remove ID30 and ID32 from the sample on the basis of their size estimates. At the best JWST resolution scales and given the modest magnification at the location of the Relic, the remaining clusters have source-plane sizes $<$100 pc. While this is significantly larger than the typical half-light radii of local GCs  \citep[$\sim$1-10 pc; e.g.,][]{Jordan2005, Spitler2006}, we expect these sources to be unresolved at the diffraction limit of JWST at $z=2.5$ (unless the magnification were to exceed $\mu>100$), motivating our requirements.

\subsection{Final Cluster Sample}
\label{sec:sample}

We detect 38 compact sources within a radius of $\sim7$ arcseconds of the Relic that pass our SNR requirements, including additional search area extended along the full extent of the tidal tail.  Given the radial and tangential magnifications, this corresponds to a search radius of roughly 50 kpc in the source plane perpendicular to the main axis ($\mu_{\mathrm{r}}=1.1$) and $\sim$40 kpc along the tidal tail axis ($\mu_{\mathrm{t}}=2.5$). The morphological analysis removes ID30 and ID32 from our final sample, which are formally resolved and thus correspond to larger star-forming complexes. 

To test for additional interlopers, photometric redshifts of all stellar sources are computed using EAzY \citep{Brammer2008}, fitting all 20 JWST bands available for each candidate with a linear combination of galaxy templates.  Following \citet{Weaver2024}, we increase the minimum error floor from the default value of 1\% to 5\% in order to more realistically reflect the calibration uncertainties of JWST/NIRCam data. We forgo the pre-processing step of iteratively tuning zero-points in order to avoid biasing our colors to those of the SED templates, also turning off both magnitude and $\beta$-slope priors for similar reasons.  All red sources are consistent with the redshift of the Relic within uncertainties.  Specifically, we mentioned earlier that ID8 has possible emission-line boosting that would support a $z=2.68$ redshift solution; however, we do not remove this cluster from our sample as the $z=2.53$ solution is still within the redshift posterior. 

While the bluest sources (see Figure~\ref{fig:sed}) have very-low-redshift solutions that are formally inconsistent ($z<0.1$), the EaZy template set cannot accommodate for higher redshift solutions for these particular objects. However, a subsequent analysis demonstrates that young simple stellar populations at $z=2.53$ can describe well these spectral shapes, and importantly are expected to exist.  We therefore do not identify any additional sources to remove from this analysis, though we do perform a series of tests to further justify the inclusion of the blue sources (see Section~\ref{sec:blue}). 

Our final sample comprises the 36 remaining point sources with sufficient SNR in F200W and F444W, all with redshift probability distribution functions and/or spectral shapes consistent with that of the Relic. The locations of these likely clusters are marked in right panel of Figure~\ref{fig:image} with white and gray circles.  

\subsection{The Origin of the Blue Clusters}
\label{sec:blue}

We note that the majority of detected clusters, especially those unambiguously in the halo, have relatively blue featureless SED shapes with no obvious emission-line boosting.  In principle, these SEDs may also be consistent with foreground cluster candidates either in the Abell~2744 intracluster medium or the far outskirts of a bright cluster galaxy. As visible in Figure~\ref{fig:image}, the nearest bright cluster galaxy is about 20 arcseconds away.  GCs associated with this galaxy would reside at a distance of around 160 kpc, not completely unrealistic but at the extremes of GC radial studies (for reference, the most distant GC in the Milky Way is located nearly 150~kpc away; \citet{Laevens14}).   

With featureless spectra and no coverage of the 4000$\mathrm{\AA}$ break or blueward at $z=0.308$, we cannot rule out the possibility that these blue clusters are unassociated with the Relic.  Deep F336W HST imaging reaching 29 ABmag, coming mid-2026 (HST-GO-14622; PIs: Whitaker, Bezanson, Leja), will provide coverage either of the rest-UV (in the case of $z=0.308$) or yield a null detection (below the Lyman limit at $z=2.53$).  
However, given their proximity to the Relic, and that they are found symmetrically around the galaxy, it is likely that they are associated with the Relic.  

We can further assess if the blue clusters in our sample are associated with the Relic based on space density arguments. We estimate the space density of clusters in the Abell 2744 by adopting a cluster catalog created following an approach similar to \citet{Harris2023}, selecting all clusters above a magnitude limit ranging from 29 ABmag to 30 ABmag. The area of the Abell 2744 is calculated to be the region where the magnification is greater than 1.5, a conservative estimate of 18.7 arcmin$^{2}$. We perform the same analysis on the Relic clusters the lie within the 7 arcsecond search radius, which corresponds to an area of 0.04 arcmin$^{2}$ in the image plane and 0.016 arcmin$^{2}$ in the source plane. We find the space density of clusters in the Relic is $\sim6-12\times$ higher, with a value of 0.29 clusters/arcsecond$^{2}$ in the Relic versus 0.03 clusters/arcsecond$^{2}$ in Abell 2744 for clusters brighter than a limit of 29.5 ABmag. Both studies would not be sensitive to detecting clusters close to the host galaxy, so this is likely a fair comparison.  If we only consider the blue clusters in this analysis (assuming the red clusters are unambiguously associated with the Relic), restricting to V-I$<$0, this overdensity decreases to a factor of $5-9\times$ higher. In either case, the space density of clusters surrounding the Relic is $>5\times$ higher relative to the Abell 2744 clusters. Moreover, the radial distribution of the clusters surrounding the Relic is also roughly similar in number to that of the Milky Way for distances of 5--10 kpc and greater \citep{Harris1996}.
We thus conclude that these clusters are likely associated with the Relic.

\subsection{Diffuse Tidal Tail Photometry}
\label{sec:diffuse}

\begin{figure*}
    \centering
\includegraphics[width=0.95\linewidth]{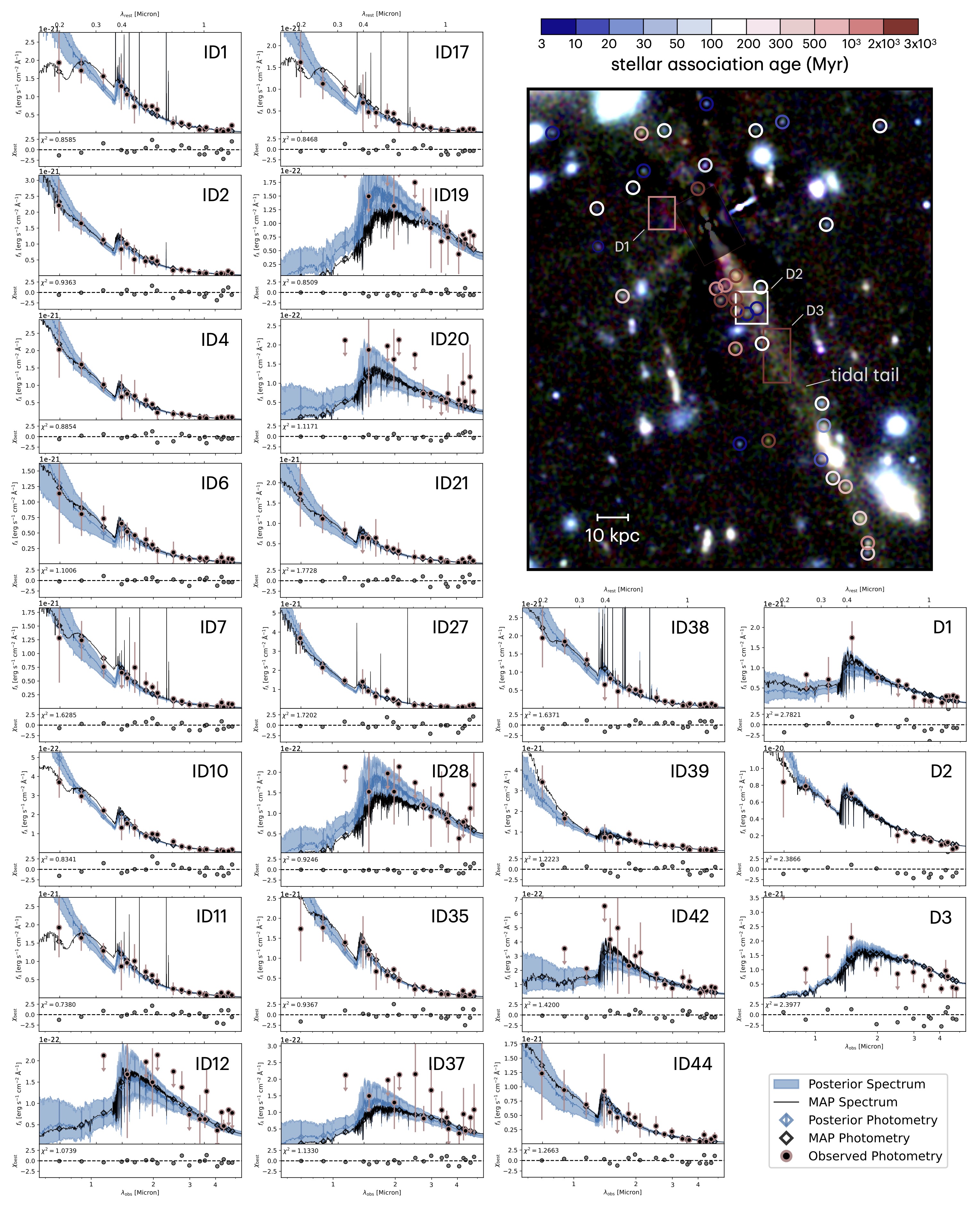}
\caption{Example SEDs of the 20 stellar clusters with reduced $\chi^{2}$ of the best fit simple stellar population is less than 1.9 at $z=2.53$. The color image (top right) shows the full sample of 36 sources, with the apertures color-coded by the maximum a posteriori (MAP) age from the single burst modeling analysis. We additionally show the extracted SED and best fit model for three regions tracing the diffuse light of tidal features (D1, D2, and D3); the box regions on the image are similarly color-coded by the best fit MAP age. The scale bar is in the observed plane, whereas the source plane scale is roughly the same perpendicular to the main axis of the Relic and a factor of 2 smaller parallel to the main axis.}
    \label{fig:sed}
\end{figure*}

In order to better understand potential systematics in the cluster photometry and search for evidence of potential biases from a dust screen (Section~\ref{sec:modeling}), we extract SEDs for three regions sampling the residual diffuse light (D1, D2, and D3; see Figure~\ref{fig:sed}). These regions are identified by eye, designed to roughly cover the width of the tidal features while also avoiding overlap with any faint blue point sources not in the segmentation map (undetected given the combined long wavelength selection). In each of these regions, we mask all detected clusters, growing the segmentation map by a factor of two, and measure the flux in each of the filters for the remaining pixels.  In all cases, we use the background-subtracted, PSF homogenized images with both contaminating neighbors and the main host galaxy light removed.  We sum the errors from the weight map in quadrature and scale up by a factor of 1.5 to account for correlated noise.  We exclude the short-wavelength medium-band filters from this analysis due to the mild residual background striping effects discussed in Section~\ref{sec:photometry}. 

Faint blue sources missing from the segmentation map may impact the SED for one particular region, D2. A few blue sources are visible in F070W but formally undetected in the redder bands -- these are likely real lower-mass (also low-SNR) young clusters not included in our analysis.  Consequently, the SED of D2 is noticeably blue in the rest-UV. Conversely, the other two diffuse regions have significantly redder SEDs, similar to those of the oldest clusters. In the following section, we model these diffuse SEDs along with the final cluster sample.

\subsection{Stellar Population Synthesis Modeling}
\label{sec:modeling}

We model the photometry of the clusters and three diffuse tidal regions using the \texttt{prospector} Bayesian inference framework \citep{prospector2021}, adopting the MIST stellar isochrones \citep{Choi2016,Dotter2016} and MILES stellar library \citep{Sanchez-Blazquez2006} from the Flexible Stellar Population Synthesis (FSPS) stellar population synthesis models \citep{fsps2009}.  Our fiducial model for all stellar sources is a simple single-burst SFH, which is generally representative of GC formation.  
For all models, we adopt a \citet{Chabrier2003} IMF and the \citet{Kriek2013} two-component dust law.  
The stellar metallicity has a top-hat prior ranging from a log(Z/Z$_{\odot}$) value of -2 to 0.2.
We also fit within the model for nebular emission (both line and continuum). The models consists of 7 free parameters (M$_{\star}$, age, dust2, logZ, dust\_ratio, dust\_index, and gas\_logU), with sampling performed using the dynamic nested sampler \texttt{dynesty} \citep{Speagle2020}. We fix the redshift to $z=2.53$ for all sources. 

Motivated by empirical results demonstrating that star clusters do not experience significant dust attenuation beyond 10 Myr \citep[e.g.,][]{Whitmore2020, Chandar2023},  we add an additional age-dust prior to the simple stellar population modeling.  We adopt a truncated normal distribution where the total dust extinction $A_{\mathrm{V}}$ exponentially drops from $A_{\mathrm{V}}=4$ for an age of 1 Myr to 0.1 by 10 Myr, with an upper bound of $A_{\mathrm{V}}=8$, a lower bound of zero, and a sigma of 3. By 10 Myr, the ionizing photon production is almost gone and feedback has essentially removed the interstellar medium \citep{Charlot00}.  The allowable range of values scales down to $A_{\mathrm{V}}$=0-0.1 mag by 10 Myr and beyond.  We tested several permutations of this dust prior, but found the results were not significantly impacted.  This age-dust prior is not used for any nonparametric SFH modeling cases tested herein, where it is no longer an appropriate assumption.

While we subtract a local background, the cluster light may still have to travel through some of this diffuse host galaxy light and may therefore be affected by dust obscuration. 
The most robust way to address this concern is through a joint spectrophotometric analysis of the host and cluster light using the IFU spectroscopy.  However, in order to test the impact of our dust assumptions on the resulting stellar masses and ages based on photometry alone, we perform a series of tests, as follows. 

First, we note that the alternative background subtraction methods described in Section~\ref{sec:photometry} yield slightly higher flux estimates.  Consequently, the stellar masses are systematically higher, with a median offset between the 50th percentile of log(stellar mass) of 0.05 dex. This is only mildly significant, with a standard deviation of $\sim$0.17 dex ($\sigma/\sqrt{N}$ of 0.03 dex).  The inferred ages show no significant systematic differences, with a median offset between of 0.02 dex and a standard deviation of 0.17 dex ($\sigma/\sqrt{N}$ of 0.03 dex).  When adding in an extra local background correction, the stellar mass and age estimates show no statistical difference relative to our fiducial adopted values.  As typical uncertainties in stellar mass are of order 0.3 dex across a similar range in ages in local studies \citep[e.g.,][]{deGrijs2005,Chandar2010}, these systematic uncertainties are not a major concern.  Moreover, the alternative approach only serves to make the clusters more massive and atypical.

Second, we re-run all fits instead allowing dust to be free.  We find that the clusters are on average 0.04 dex more massive, 300 Myr younger, with an average $A_{\mathrm{V}}$ of 0.4 mag.  By comparison, the main host galaxy central diffuse light has an inferred A$_{V}$ value of 0.2$^{+0.3}_{-0.12}$ mag, which is lower than average at these redshifts \citep[e.g.,][]{Siegel2025}.  Regardless, the fits still return solutions for old clusters and the overall trends remain the same.  Given that our subsequent interpretation of the formation history of the Relic does not significantly change when allowing dust to be free in the modeling analysis, we defer a more comprehensive dust analysis to future work that includes the IFU spectroscopy.

Third, we generate a series of SEDs by extracting the diffuse residual light of the tidal features in a few representative regions (Section~\ref{sec:diffuse}).  We model these diffuse SEDs using the same assumptions as above.  The goal of this exercise is to constrain the shape of the diffuse SED (age, dust) in order to test the impact.  When adopting the same age-dust prior for a simple stellar population, we find that two of the three regions have old ages broadly consistent with the main host galaxy, with an age of 1.96$^{+0.41}_{-0.44}$ Gyr in D1 and 1.13$^{+1.00}_{-0.56}$ Gyr in D3. However, region D2 is a notable outlier, with a significantly bluer SED and age of 120$\pm$20 Myr. We also run modeling with dust attenuation free, adopting a nonparametric SFH as done when modeling the main galaxy, in order to test if the clusters may be sitting behind an extra dust attenuation screen different from their intrinsic properties. While we recover dust attenuation values ranging form 0.3 to 1 mag, these fits are notably worse than those assuming a simple stellar population, with the $\chi^2$ a factor of two higher on average.  Altogether, while we cannot rule out that a more complex dust geometry is at play, the effect is overall subtle and thus unlikely to impact the inferred stellar population modeling results.  Future work with the JWST/IFU spectroscopy will further illuminate this issue.

We split the results of our modeling analysis into two groups based on goodness of fit.  We find 21 clusters with a reduced $\chi^2<2$; this includes ID30 and ID32, which did not make it into the final sample based on morphology (Section~\ref{sec:sizes}).  The remaining 15 clusters have reduced $\chi^2>2$ (grey circles, Figure~\ref{fig:image}).  Though adopting an alternative background photometry method (Section~\ref{sec:photometry}) adds 6 additional clusters (ID3, ID9, ID15, ID18, ID24, and ID31), these sources largely have low SNRs and more uncertain background subtractions; we note their location with teal circles in Figure~\ref{fig:image}, but do not include them hereafter.  We summarize the results of our stellar population modeling in the following section, only including the 36 robust clusters that comprise the final sample hereafter.

\begin{figure}
    \centering
\includegraphics[width=\linewidth]{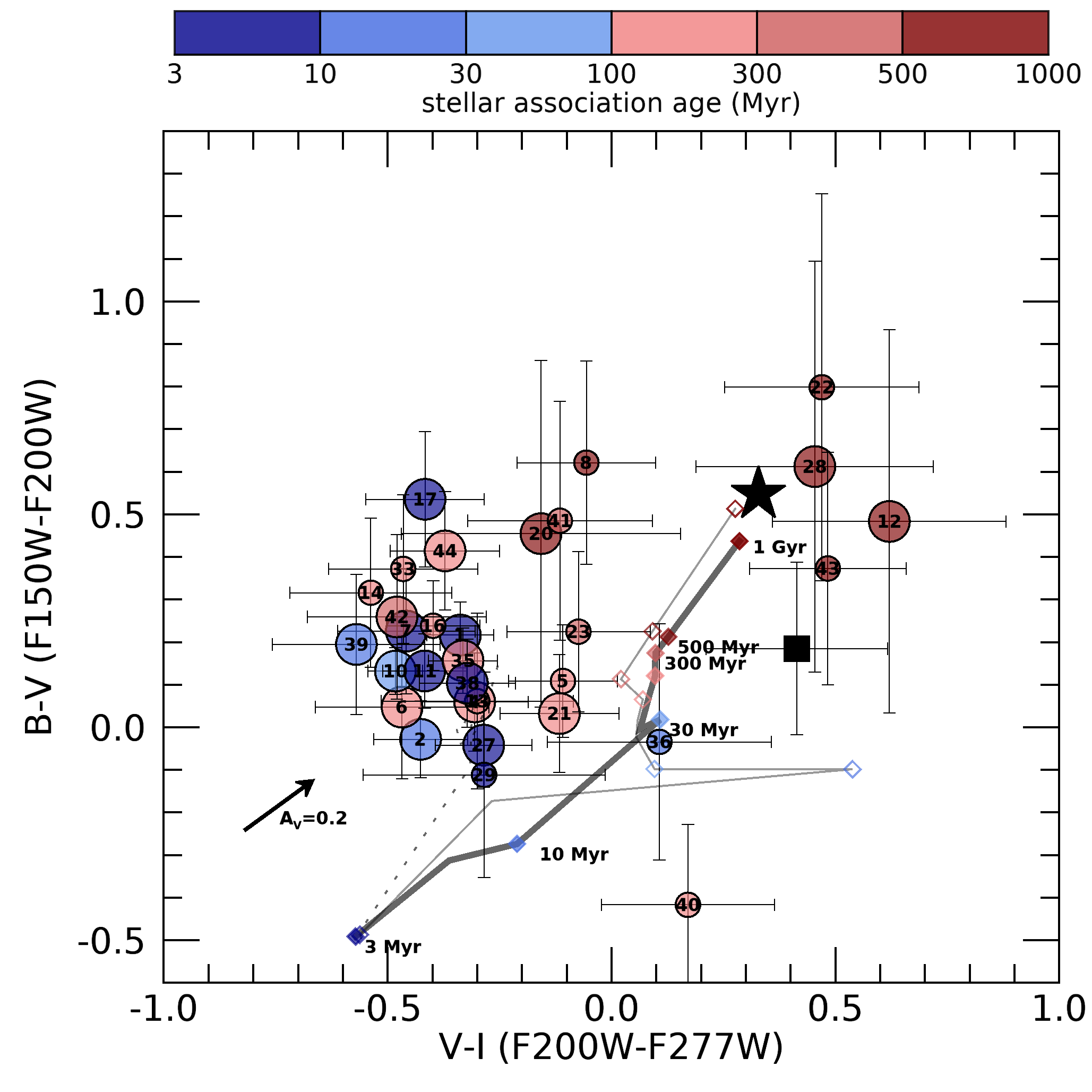}
\caption{Pseudo B-V versus V-I rest-frame color-color diagram through the F150W, F200W, and F277W filters for all compact sources detected near the Relic and the two neighboring galaxies along the tidal tail (see Figure~\ref{fig:neighbors})  The symbols are color-coded by the age of their best fit simple stellar population model, where larger symbols represent fits with reduced $\chi^{2}<2$ and smaller circles are for reduced $\chi^{2}>2$. The color of the main Relic is indicated with a star and diffuse light in the tidal tail with a square symbol.  The arrow represents the typical dust attenuation for ages less than 10 Myr of 0.2 mag, assuming the median dust index value of -0.9. FSPS model tracks for 25\% solar metallicity at $z=2.53$ are shown (solid line), with solar metallicity indicated with a thinner higher transparency line.  The 25\% solar metallicity FSPS model track for the same three filters but now at $z=0.308$, the redshift of the Abell 2744 cluster, is shown with a dotted line. }
    \label{fig:color}
\end{figure}

\section{Results}\label{sec:results}

Figure~\ref{fig:color} shows the observed color-color diagram for the Relic clusters in filters roughly equivalent to the rest-frame B-V (F150W-F200W) and V-I (F200W-F277W) at $z=2.5$.  While in principle the added coverage of the medium-band filters is useful, these images are not as deep and the SW filters have noticeable residual background effects. For this reason, we only show a color-color diagram using the deeper broadband filters. Comparing these broadband colors via aperture photometry to both the model tracks from \citet{bc03} and our inferred (color-coded) stellar association ages is a logical sanity check of our more complex modeling efforts and assumptions.  The thick model line represents a lower metallicity, at 25\% of the solar value, whereas the thin line is for models with solar metallicity.  We find three broad groups of clumps in color space: those with blue colors and ages predominantly in the range of 8-20 Myr, an intermediate group with ages ranging from 100-600 Myr, and an older population with ages $>$1 Gyr, close to the mass-weighted age of the Relic itself (star symbol) and the diffuse light of the tidal tail (square symbol).  Among the clumps younger than 10 Myr, the typical dust attenuation is 0.2 mag, with a preferred low dust index value of -0.9 (see arrow in Figure~\ref{fig:color}).  As is often the case, the intrinsically red clumps are faint and thus carry the largest uncertainties.  The one outlier with red V-I and blue B-V colors (ID40) is located within a few kpc of the tidal tail with an SED that shows a clear break marked mostly by upper limits in the rest-UV and a highly uncertain B-band magnitude that contributes to the atypical colors on this diagram.  These uncertainties aside, we find that the best fit ages and observed colors are broadly consistent. 

The SEDs for 21 out of the 22 clusters with best fit model statistics of reduced $\chi^{2}<2$ are shown in Figure~\ref{fig:sed} (we exclude ID34 for space purposes only, which has $\chi^{2}=1.9$). The blue shaded region in each SED shows the 16th, 50th and 84th percentile ranges.  The maximum a posteriori (MAP) spectrum for the youngest clusters is often outside of the full 16th-84th distribution.  Further inspection of the age distributions reveals asymmetry, with a roughly one-sided Gaussian peaked at ages of 8-10 Myr.  For all of these cases we adopt the best fit MAP age as our fiducial and adjust the upper limit to be the 50th percentile.  

\begin{figure}
    \centering
\includegraphics[width=\linewidth]{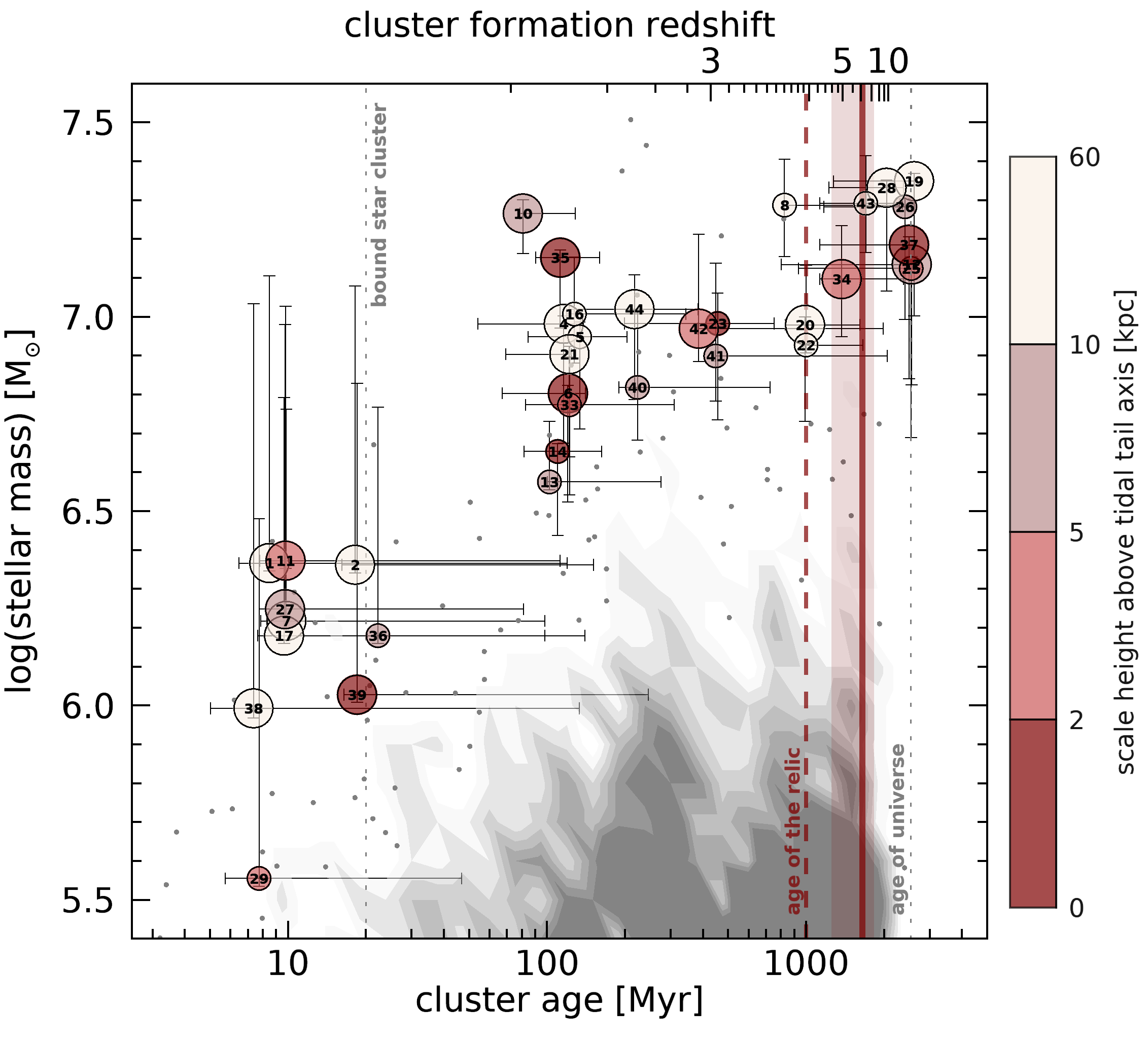}
\caption{Clusters near the Relic are generally more massive and older. The color-coding represents the scale height above the tidal tail axis (which is close to the source plane value given $\mu_{\mathrm{r}}=1.10$), with the tidal tail extending roughly 100 kpc from the center of the Relic (or 40 kpc in the source plane, with $\mu_{\mathrm{t}}=2.49$).  The average ages of the Relic (vertical maroon lines) when adopting either a nonparametric SFH (solid line) or a simple stellar population (dashed line) are shown for reference. The top axis shows the cluster formation redshifts relative to the age of the Universe at $z=2.53$ (dotted line). The Relic clusters are consistent with the extremes of model predictions at $z=2.5$ from \citet{Pfeffer2024} (greyscale contour and individual grey points for outliers; see Section~\ref{sec:theory} for more details). }
    \label{fig:age_mass}
\end{figure}

The top-right false-color image in Figure~\ref{fig:sed} shows the final sample of 36 stellar sources, with the color of each circle represented by the age of the MAP solutions.  In general, clusters unambiguously in the halo have younger ages in the range of 8-200 Myr, and those embedded within the tidal tail are generally older with ages $>1$ Gyr.  That said, there are a few $\sim$8 Myr clusters nearby to these oldest clusters. This can also be seen more explicitly in Figure~\ref{fig:age_mass}, as described next.

\begin{figure}
    \centering
\includegraphics[width=\linewidth]{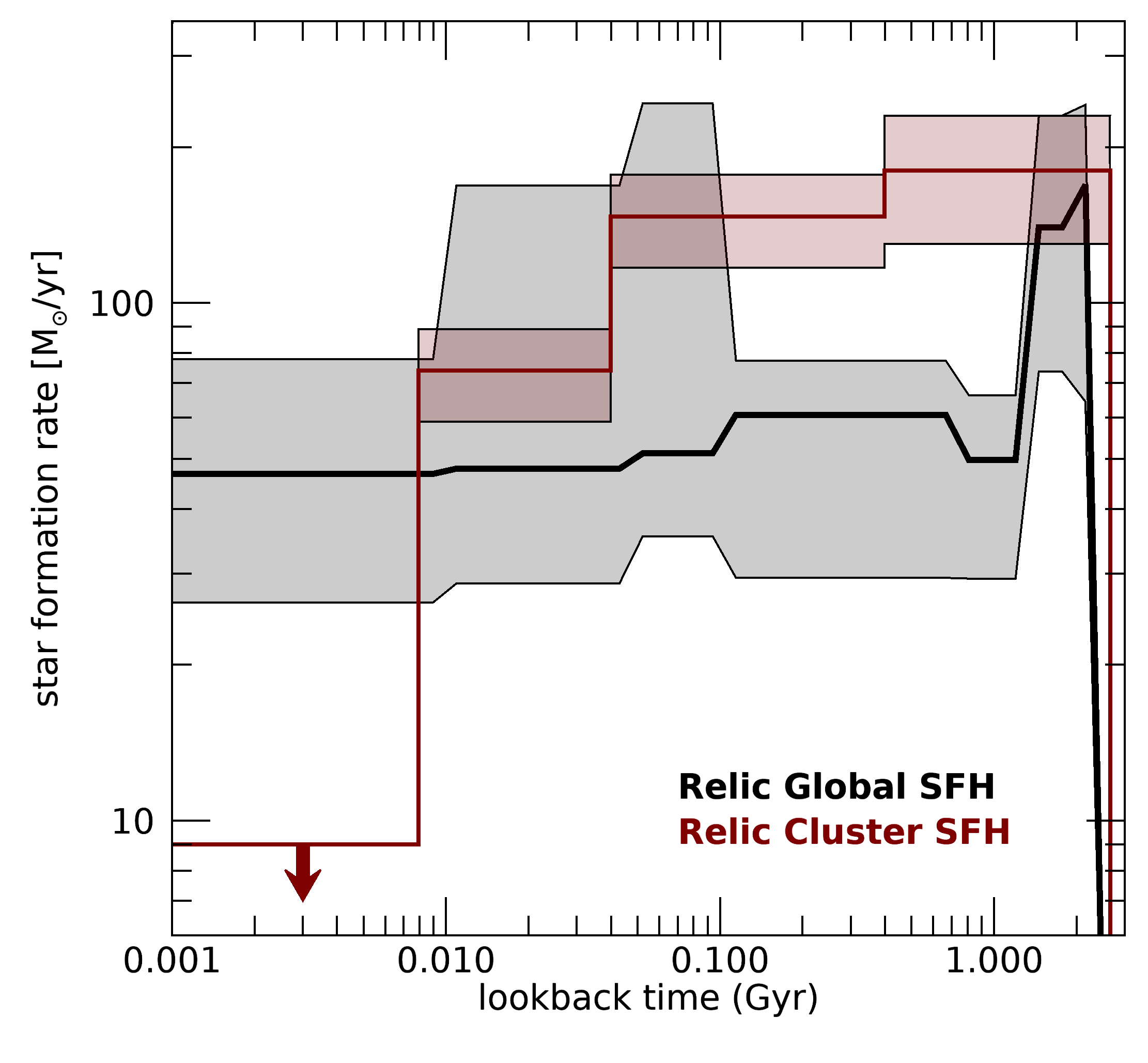}
\caption{The star formation history of the main galaxy (black) is broadly similar to the cluster-inferred SFH (maroon).  The cluster-implied SFR is $\sim$2-3$\times$ higher for lookback times $0.1-1$ Gyr.  As there are no clusters less than 8 Myr, we show a 3$\sigma$ upper limit. Our favored interpretation is that this excess supports a scenario where some fraction of the young and intermediate age clusters were accreted from the rich overdensity the Relic resides within. }
    \label{fig:sfh}
\end{figure}

We next consider the physical properties of the clusters in Figure~\ref{fig:age_mass}.  We see a clear trend where older clusters have systematically larger stellar masses. As the data are not sensitive to faded, lower-mass older clusters, this trend is effectively a consequence of incompleteness.  Each symbol is color-coded by the projected scale height above the tidal tail axis. The majority of older clusters lie within a few kpc scale height of the tidal tail main axis, engulfed within the diffuse light, with many close to the main galaxy (not in the halo).  The estimated age of the Relic itself is shown as a vertical line. The solid line represents the mass-weighted age of 1.65$^{+0.40}_{-0.18}$~Gyr when modeling the photometry released in \citet{Suess2024}  (Figure~\ref{fig:neighbors}) with a nonparametric SFH (Figure~\ref{fig:sfh}) that adopts a mass-metallicity prior from \citet{Gallazzi2005}.  We also fit the same photometry with a single burst SFH (dashed line in Figure~\ref{fig:age_mass}), but note that this fit cannot adequately describe the full SED shape, either finding a younger poststarburst age that overpredicts the strength of the Balmer-break (the solution shown in Figure~\ref{fig:age_mass}), or an older age consistent with the nonparametric SFH best fit that cannot describe the rest-UV light well. In either case, the Relic is consistent overall with a moderately old stellar population, older than the majority of the associated clusters.

From the top axis of Figure~\ref{fig:age_mass}, we see the formation redshift inferred for each cluster.  This ranges from the epoch of observation at $z=2.5$ for the youngest clusters up to $z>10$ for the oldest clusters.  The majority of clusters were formed at $z=2.7$ or earlier.  Note that the age of the Universe at $z=2.53$ is only 2.5 Gyr, or a lookback time of almost 11 Gyr.  The ages of nearby GCs range between roughly 11 and 13 Gyr \cite[e.g.,][]{DeAngeli2005,MarinFranch2009,Ying23,Ying24}.  While these dense star clusters are often thought of as some of the oldest structures in the Universe, their formation redshifts likely includes $z=2.5$ (where we observe the Relic) and earlier.

\section{Discussion}\label{sec:discussion}

\subsection{Cluster Population in the Relic}
\label{sec:cluster_overview}
The discovery of the Relic at $z=2.53$ presents a unique opportunity to establish the first direct connection between the formation of long-lived, stellar clusters at high redshift and local stellar populations.  At the peak epoch of star formation, this massive quiescent galaxy exists at a time that bookends the era of GC formation, with cluster ages ranging from less than 10 Myr up to 2 Gyr.  If we were to fast-forward these stellar associations to the present day, their ages would range from 11 to 13 billion years, should they survive disruption (more below).

The clusters have estimated masses of $\approx10^6~M_{\odot}$ up to $\approx \mbox{few}\times10^7~M_{\odot}$.
This overlaps with local samples of interacting and post-merger galaxies which have formed clusters with similar ages and with similarly high masses, of order 10$^{7}$-10$^{8}$ M$_{\odot}$, including the Antennae \citep{He22}, NGC~34 \citep{Schweizer07, Adamo2020b}, and NGC~7252 \citep{Schweizer98}.
Note however, that clumps discovered in other more highly magnified systems have estimated half-light radii ranging from $\approx20$~pc down to 1~pc \citep[e.g., the Sunburst arc at z$=$2.37 and Sunrise Arc at z$\sim$6;][]{Vanzella2022a, Vanzella2023}. These smaller sizes are very similar to those measured for GCs in nearby elliptical and spiral galaxies including the Milky Way \citep[e.g., $<$1~pc to 8~pc;][]{Kundu1998, Kundu1999, Larsen2001, Jordan2005, Spitler2006}.
Taken together, the Relic presents a compelling case where we are observing bona fide GCs soon after they formed, as evidenced by their derived ages, stellar masses, and (unresolved) sizes.

The oldest clusters, which also have the largest stellar masses, only survive such long timescales if they are gravitationally bound.  It is expected that these clusters will continue to shed mass through two-body interactions as they evolve to the present day. 
Simulations predict up to an order-of-magnitude loss in stellar mass \citep[e.g.][]{Pfeffer2024}, representing a trajectory in the age-mass plane that could naturally evolve the oldest Relic clusters to the present-day GC population.   
Given their high masses, it is therefore logical to speculate that the GCs in the Relic will survive evaporation driven by two-body relaxation for the age of the Universe.  The \citet{Pfeffer2024} simulations presented in Figure~\ref{fig:age_mass} also nicely demonstrate how we are only seeing the `tip of the iceberg', with the bulk of cluster population well below our detection limit.

While spectroscopy is needed to robustly constrain metallicity, our photometric analysis tentatively suggests that the older clusters have higher metallicities of order log(Z/Z$_{\odot}$)$\sim-0.5$, whereas the younger clusters have metallicities $\sim1$ dex lower with log(Z/Z$_{\odot}$)$\sim-1.5$. 
The typical posterior ranges up to 0.5 dex in width for ages less than 500 Myr, but is a factor of two larger for older clusters where the age-metallicity degeneracy hinders interpretation. Many clusters have metallicities closer to the 25\% solar metallicity trends relative to solar metallicity, as shown in Figure~\ref{fig:color}, though there isn't much distinguishing power with these particular colors. 
In any case, both the young and old populations have inferred metallicities that would largely be considered ``Population I'' metal-rich clusters according to theoretical predictions \citep[e.g.,][]{Chen2024}.  
Given that the Relic resides in a massive dark matter halo, it is natural to expect the GCs to be dominated by Pop-I populations that formed through shock compression due to gravitational collapse, such as mergers \citep{Chen2024}. However, the metallicity prior does not extend below -2 and thus our analysis is not sensitive to recovering metal-poor clusters by not allowing for such solutions.  Furthermore, in the absence of spectroscopy, these metallicity measurements should be interpreted with caution. JWST/IFU spectroscopy from JWST-GO-6405 (PIs: Cutler/Whitaker) will provide better constraints on metallicity in order to distinguish formation scenarios. 
We also note that we are not sensitive to detecting stellar sources in the core of the Relic, where they may have higher metallicities.  The central diffuse light of the host galaxy is overall metal-rich when adopting a nonparametric SFH and a mass-metallicity prior following \citet{Gallazzi2005}, with log(Z/Z$_{\odot})=0.07{+0.09}_{-0.17}$.

The Relic exists in a known overdensity \citep{Naidu2024}, which laid the groundwork with a rich history prior to the epoch of observation.  The Relic environment likely fueled past episodes of in situ star formation and increased the probability of accretion events and dynamical interactions.  
The age and mass distributions of cluster systems in local galaxies are known to closely trace the past major star formation episodes, correlating with inferred star formation histories of the host galaxies \citep[e.g.,][]{Chandar17, Chandar21}.  In the case of the Relic, the clusters suggest three episodes of star formation (see Figure~\ref{fig:age_mass}).  If we assume a typical space velocity of 700 km/s for a galaxy in a proto-cluster and adopt a projected distance of 70 kpc to the nearest companion, this suggests the Relic flew by roughly 100 Myrs ago.  In this context, the intermediate-age population of stellar clusters may have formed at least in part as the result of this particular interaction (more discussion in Section~\ref{sec:sfh}). In principle, we would expect to see a similar rejuvenation in the host galaxy light; however, the uncertainties in the SFH are too large to reach a definitely conclusion (Figure~\ref{fig:sfh}). On the other hand, the oldest clusters are consistent with the mass-weighted age of the Relic, likely forming in tandem with the main stellar population.  

\subsection{Formation History of the Relic and its Clusters}
\label{sec:sfh}

While our ability to draw inferences directly from a photometry-based SFH are limited, there is suggestive evidence from the diffuse host galaxy light that somewhat tells the same story as the cluster population.  As described in Section~\ref{sec:photometry}, the SED of the main galaxy represents the spectral shape within the inner 0.5 arcseconds but is corrected to total through aperture corrections.  This has become a standard approach in the literature.  As the light is completely distinct from the clump population, the resulting SFH serves as an interesting comparison point.  

The SFH of the Relic in Figure~\ref{fig:sfh} shows a main epoch of star formation 1-2 Gyr ago, with a relatively flat SFH thereafter, with a star formation rate of 50 M$_{\odot}$ yr$^{-1}$.  An enhancement coinciding with the dynamical interaction with the two neighbors $\sim$100 Myr in the past is only hinted at via the larger uncertainties at this epoch.
For comparison, we also estimate a cluster-based SFH.  In short, we use the stellar masses of the clusters in the different age intervals together with observed local star formation rate (SFR)- cluster mass scaling relations to infer the SFR associated with each cluster \citep{Chandar21}.  This is then integrated based on all of the clumps within that age (lookback time) bin, resulting in a cluster-inferred SFH (maroon line in Figure~\ref{fig:sfh}).  The caveat here is the unknown applicability of local scaling relations at high redshift -- we expect significant progress in this area in the coming decade.  In the following paragraphs, we discuss the similarities and differences between these two SFHs in the context of the observations as a whole in order to piece together the possible story of the Relic.  

The formation of young clusters may be a natural outcome in the aftermath of the dynamical interaction if the Relic accreted new cold gas. Alternatively, the moderate rest-UV light of the host galaxy also hints that they can form in tandem with the main galaxy owing to the ongoing lower-level star formation. Consistent with the former interpretation, the cluster distribution in the Relic shows a striking similarity to that of the local post-merger NGC~34 \citep{Adamo2020}. Foremost, NGC~34 hosts similarly seemingly ``too-massive'' GCs (see also Section~\ref{sec:toomassive}). Moreover, the spatial distribution of GC ages is quite similar between the two galaxies.  Thus, we speculate that the young halo clusters formed from tidally stripped gas from the neighboring galaxies that was thrown out into the halo in the wake of these systems, as observed in NGC~34.

However, there exists a discrepancy between the cluster- and galaxy-inferred SFHs. Moreover, it is an open question as to how clusters get out into the halo. It has been suggested that clusters can form in a protodisk and then be ejected into the halo due to violent relaxation \citep{Toomre72}.  While mergers can drive this violent relaxation, galaxies that experience such a recent merger would generally also show morphological disturbances \citep[e.g.,][]{Toomre72,Barnes98,Mihos94,Goudfrooij07} rather than presenting as a smooth elliptical galaxy.  While there does exist a hint of a red tidal arc (see Figure~\ref{fig:image}) and a prominent stream of stars in the wake of the Relic, there is otherwise no obvious evidence for a recent major merger.  Moreover, we do not see compelling evidence for a disk-like structure in the Relic. That said, we cannot rule out the presence of a disk from morphology alone given the existence of fast-rotating early-type galaxies at both $z\sim0$ \citep{Cappellari2016, Graham2018} and $z\sim2$ \citep{Newman2018}.  

Instead, it may be that the young to intermediate-age clusters hold the best clues to the Relic's formation history: despite the systemic quenching of the galaxy roughly one billion years prior (black line, Figure~\ref{fig:sfh}), it appears that clusters continued to form.  This robust cluster population indicates an SFR that is $\approx2-3$ times higher than the integrated light at lookback times of 0.1-1 Gyr.  So while it is logical to infer that the clusters closest to the tidal tail formed as a result of the past dynamical interaction, it is harder to explain halo clusters with the youngest ages.  These $\sim10-100$~Myr clusters, shown in the center panel of Figure~\ref{fig:sed} as white circles, are distributed throughout the halo of the Relic. Whereas the ejection scenario described above does not seem plausible, some fraction may have logically formed in stripped tidal streams of gas.  Another possible explanation is that some of these clusters instead were accreted from smaller, nearby galaxies, which are common in the dense environment surrounding the Relic.  
A number of the halo GCs in the Milky Way have been accreted in this way \citep[e.g.,][]{BrodieStrader2006,Forbes2010,Ishchenko2023,Belokurov2024}.  Moreover, local elliptical galaxies in dense environments are believed to accrete a significant fraction of their GC populations \citep[e.g.,][]{BrodieStrader2006}. 

The lower metallicities of these clusters, be that from metal-poor dwarf galaxies or metal-poor stripped gas, only serve to further bolster this hypothesis (see Section~\ref{sec:cluster_overview}).  It is therefore the notable gap at lookback times less than 1 Gyr
between the cluster-based and main galaxy SFH together with the spatial distribution and lower metallicities of these young to intermediate age clusters that builds the case for an accretion origin.  
While speculative, this accretion interpretation also fits nicely into the larger hierarchical framework of galaxy formation, where the outskirts of quiescent galaxies show empirical evidence of significant growth over their subsequent many billions of years of evolution \citep[e.g.,][]{vanDokkum2010,Hill2017}.  While the bulk of star formation happened in the distant past in the Relic, it will continue to evolve through minor mergers and accretion -- tentatively supported herein by the existence of these young and intermediate-age halo clusters.

All together, we see evidence for an age spread of $\sim$2 Gyr for the GCs in this massive elliptical galaxy. While we present possible interpretations above, photometric uncertainties preclude a more definitive answer as to if these clusters formed in different bursts, more continuously, and/or if some fraction were accreted. The follow-up ultra-deep spectroscopic data (JWST-GO-2561) will provide more information regarding the true formation history of the clusters within the Relic. 

\subsection{Implications for the Most-massive Clusters}
\label{sec:toomassive}

While we note some local observations of ``too-massive'' GCs in Section~\ref{sec:sfh}, it is rare for stellar masses to exceed $>10^{6-7}$ M$_{\odot}$ locally; yet, it is quite common at higher redshifts \citep[e.g.,][]{Claeyssens2026}. An order-of-magnitude baseline for what to expect at high redshift comes from constraints of the Milky Way stellar halo composition, where 11$\pm$1\% of the halo is estimated to be disrupted GCs \citep[$\sim$10$^{8}$ M$_{\sun}$;][]{Koch2019}.  As a simple thought experiment, we consider all clusters with log(M$_{\star}$/M$_{\odot}$)$>$6.5 and assume that 27\% will disband by $z=0$, while the rest experience 0.18 dex in mass loss (the median equivalent value at these masses in the \citealt{Pfeffer2024} models, see Section~\ref{sec:theory}). With a combined stellar mass of $\sim$10$^{8.5}$ M$_{\sun}$ for GCs beyond $\sim$4 kpc, the halo of the Relic would eventually subsume roughly $\sim$10$^{8.4}$ M$_{\sun}$ for the disbanded clusters and, less important, $\sim$10$^{7.0}$ M$_{\sun}$ from mass loss.

However, the Relic is significantly more massive than the Milky Way already at $z=2.5$.  If one tracks the main progenitor of the Milky Way through an empirical number-density–based progenitor–descendant matching approach back to $z\sim2.5$, it would have roughly 10\% of the present day stellar mass \citep{vanDokkum2013}. A simple linear scaling by the ratio of the Relic stellar mass to that of the Milky Way at $z\sim2.5$ implies that the Relic will host 10$^{9.3}$ M$_{\odot}$ in disrupted GCs by $z=0$. This number may be even higher given that the most massive local galaxies are known to host higher specific frequencies of clusters \citep{Peng08}.  Although we are not sensitive to detecting GCs closer in to the host galaxy (within $\sim$4 kpc), and thus there is a missing correction factor here, our high inferred stellar masses do not collectively appear to cause significant tension.

More broadly, it is worth noting that  there may be alternative explanations for the large stellar masses of the oldest clusters that are worth exploring in future work. They may not be GCs at all, but rather ultra-compact dwarf galaxies that form through a previously unrecognized formation channel. Or perhaps the Relic is one data point in a growing body of evidence \citep[e.g.,][]{Claeyssens2026} that the GC mass function (GCMF) at early cosmic times may be skewed toward higher masses than observed locally \citep[e.g., some local examples in dwarf galaxies include][]{vanDokkum2018, Li2024}. While no single system yet requires invoking a top-heavy initial GCMF, taken cumulatively such observations may point to modest tension with simple extrapolations of present-day cluster populations. It is common to invoke substantial mass loss between $z=2.5$ at $z=0$ as a solution, but there may be something else at play in these high-redshift galaxies yet to be understood.

\subsection{Comparison with Theoretical Predictions}
\label{sec:theory}

\begin{figure*}[ht]
    \centering
    \includegraphics[width=0.45\textwidth]{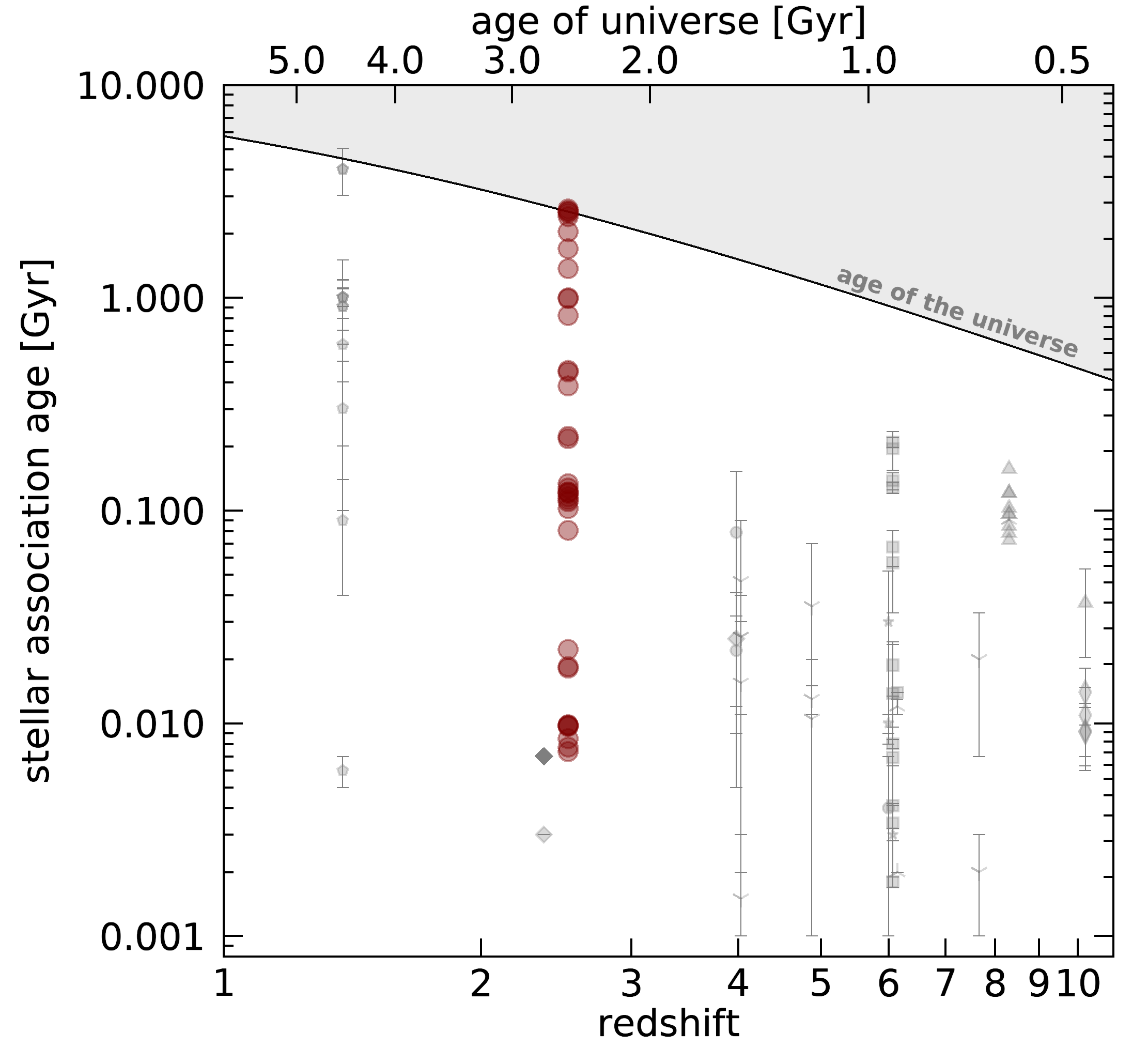}
    \label{fig:mod1}
    \includegraphics[width=0.45\textwidth]{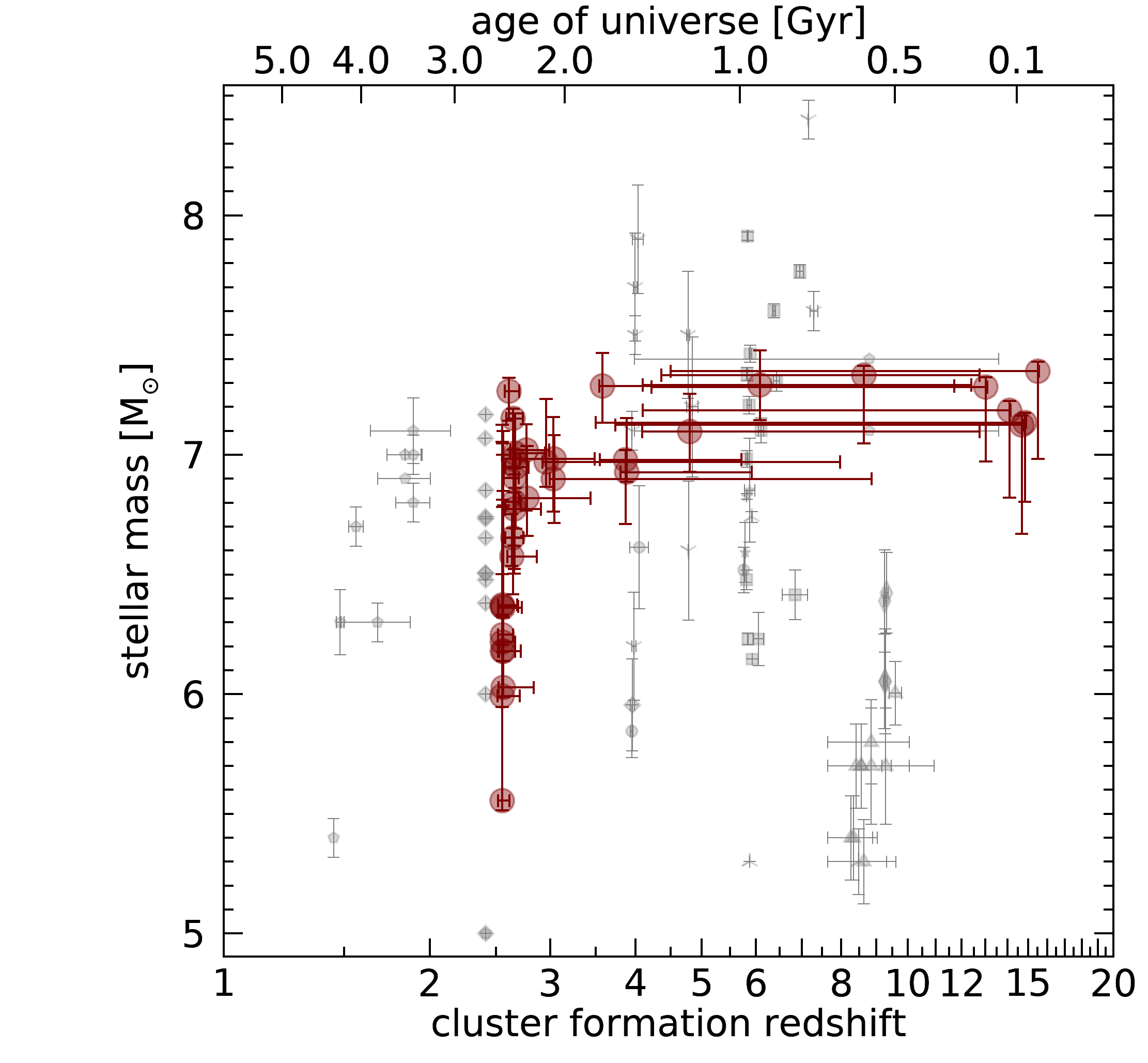} \label{fig:mod2}
    \caption{Compilation of mass, age, and associated formation redshift estimates for compact stellar sources discovered in lensed high-redshift galaxies with JWST photometry from \citet{Pfeffer2024}: Cosmic Gems \citep[thin diamond;][]{Adamo2024}, Firefly Sparkle \citep[triangle;][]{Mowla2024}, MACS0416 \citep[upward triangle;][]{Messa2024a}, Sunrise \citep[star;][]{Vanzella2023}, Cosmic Grapes \citep[square;][]{Fujimoto2024}, A2744 \citep[circle;][]{Vanzella2022b}, Sparkler \citep[pentagon;][]{Mowla2022}, RCS0224 and MACS0940 \citep[downward triangle;][]{Messa2024b}.  Plus one additional set of data based on HST photometry: Sunburst \citep[diamond;][]{Vanzella2022a}.  The Relic cluster age, formation redshift, and mass estimates are shown as maroon circles for comparison.}
    \label{fig:compilation}
\end{figure*}

The E-MOSAICS project (MOdelling Star cluster population Assembly In Cosmological Simulations within EAGLE) offers cosmological hydrodynamical simulations of galaxy formation that allow for the formation and evolution of star clusters using subgrid models \citep{Pfeffer2018,Kruijssen2019}.  Following their formation, clusters in the simulation lose mass due to stellar evolution, two-body relaxation, and shocks from rapidly changing tidal fields.  Both young and old star clusters are assumed to form and evolve following the same physical mechanisms.
The predicted ages and masses of clusters from the fiducial model of \citet{Pfeffer2024} at $z=2.5$ are shown in grayscale in Figure~\ref{fig:age_mass}, with the most extreme cluster masses (at different ages) plotted as the small gray circles.  Our mass and age estimates for clusters in the Relic fall at the extreme upper end of the model predictions at a redshift of $z=2.5$. 

While significant mass loss may not be necessary given the high-mass clusters found in nearby post-merger systems \citep[e.g.,][]{Schweizer98, Schweizer07, He22}, we can make a very rough estimate of the expected mass loss following \citet{Pozzetti2019}, where the number of bright clusters at high redshift directly scales with the implied mass loss.  As we detect 25 clusters brighter than 30th magnitude in F444W, we might expect a subsequent mass loss of around 0.2 dex per cluster by $z=0$.  This number compares well to the results of \citet{Pfeffer2024}, where 70\% (73\%) of all clusters (log(M$_{\star}$/M$_{\odot}>6.5$) survive disruption. These models predict a slightly larger median stellar mass loss of 0.27 dex (0.18 dex) from $z=2.5$ to $z=0$, though consistent within the normalized median absolute deviation of 0.08 dex (0.05 dex).  That said, there is not yet a broad consensus among GC formation models regarding the fraction that survive \citep[e.g.,][]{Valenzuela2025}.

\subsection{Comparison with High-redshift Cluster Observations}

 In Figure~\ref{fig:mod1}, we compare the properties of the clusters in the Relic with those determined for other high-redshift, lensed galaxies.  Figure~\ref{fig:mod1}a shows the age of each cluster with redshift, with the literature compilation of observed stellar sources presented in \citet{Pfeffer2024}.  The Relic is shown by the maroon circles.  Here we see that the ages of the clusters in the Relic are among the oldest identified thus far for any high-redshift galaxy.  Figure~\ref{fig:mod2}b shows the formation redshift of the clusters relative to their stellar mass.  While the Relic includes a relatively older population of clusters with earlier formation redshifts, the general range of stellar masses is similar to other measurements at high redshift.  However, the Relic is unique among these galaxies in that it is an early-type galaxy, whereas the others are clearly later-type systems with ongoing star formation.  Taken at face value, this indicates that elliptical galaxies started to form stars and clusters earlier than most gas-rich disk galaxies. This is also consistent with the dense environment where the Relic resides and the downsizing paradigm of galaxy formation as a whole \citep{Thomas2005,Thomas2010}.  

\subsection{Link to Exotic Populations}

Beyond their relationship to galaxy assembly, gravitationally bound, compact collections of stars that comprise GCs are also known to produce exotic stellar populations \citep{Gratton2019}.  Dynamical interactions within GCs give rise to exotic systems, including X-ray binaries, pulsars, fast radio bursts, and merging BH systems \citep{Giesler2018, Ivanova2008}.  The rate of binary black hole (BBH) mergers originating from GCs carries information about their abundance in the early Universe, as well as their mass and radius distributions.  It is thought that more-massive, compact GCs give rise to more BBHs \citep{Fishbach2023}.  While current GW detectors only reach $z\sim1$, planned upgrades will extend detection to $z>2$ within the next few years \citep{Evans2021, Kalogera2019}. If candidate massive clusters identified by JWST at high redshift were to survive a Hubble time, this would make them the progenitors of the local, metal-poor GCs that are factories for BBH mergers.  While most clusters in the Relic won't be old enough yet, the most-massive, metal-poor clusters may already be marking the first sites of BBH production, with GW signatures detectable in the near future.

\section{Conclusions}

In this paper, we overview the physical properties of 36 compact stellar sources, likely star clusters,  associated with a massive, quiescent galaxy at $z=2.53$ behind the Abell~2744 galaxy cluster. The clusters tell a story of the formation history of the Relic that provides a unique laboratory for testing globular cluster formation theories.  Below we summarize the main results,

\begin{itemize}
    \item The stellar populations of the Relic clusters comprise three main age groups.  The oldest stars $>1$ Gyr likely formed in situ at $z>5$ in tandem with the bulk of the host galaxy.  The intermediate-age population, with ages of $\sim$100-500 Myr, are thought to be the result of both a tidal interaction with two nearby low-mass quiescent galaxies as well as clusters formed ex situ and acquired via accretion events.  While the galaxy is overall quenched, we see clear evidence for a young cluster population (10-100 Myr) that exists system-wide spatially and is also suspected to have been accreted. Altogether, if the 36 clumps survive to the present day, they bookend the typical age range of local globular cluster systems of 11-13 Gyr. 
    \item The Relic hosts over a dozen clusters that are sufficiently old to be considered both gravitationally bound and likely to survive subsequent mass loss due to dynamic interactions.  These clusters are the best known examples of globular cluster candidates at high redshift, given the combination of their old ages, compact/unresolved sizes, and moderately high stellar masses.  Moreover, they are the oldest clusters observed to date at similar redshifts, and unique in that their host galaxy is globally quenched already by $z=2.5$.
    \item The stellar masses of the oldest clusters are of order log(M/M$_{\odot}$)$\sim$7, consistent with the largest formed globular clusters in model predictions. Over the subsequent 11 Gyr to the present-day, it is thought that some fraction of these compact, gravitationally-bound stellar systems will be disrupted, whereas the rest will experience up to an order-of-magnitude loss in stellar mass due to dynamical interactions.
    
\end{itemize}
 
The overdensity within which the Relic resides was discovered in one of the deepest, most homogeneous JWST datasets to date.  This data will continue to serve as a fertile playground to search for clusters surrounding a wide range in host galaxy type and mass \citep[e.g., see][]{Claeyssens2024}, presenting a unique opportunity to test globular cluster formation scenarios.  In particular, globular clusters exist with high specific frequencies in both the lowest- and highest-mass quenched galaxies in the local Universe \citep{Olsen2004,Peng08}. These massive quiescent galaxies form rapidly and shut down as early as $z\sim5$ \citep{Carnall2023, deGraaff2024, Kakimoto2024, AntwiDanso2025}, whereas at least some fraction of the lower-mass quiescent population quenches at slightly later times closer to $z\sim1-2$ \citep{Cutler2024}.  In either case, studies of the globular cluster populations within these galaxies need to push to $z>2$ in order to distinguish in situ formation versus accretion scenarios. Studies of rich overdensities of quiescent galaxies at cosmic noon will therefore enable us to draw evolutionary connections with globular cluster formation theories for larger samples.  Moreover, there exist even deeper programs (e.g., JWST-GO-3293, PIs: Atek/Chisholm), where studies of the globular cluster luminosity function can be pushed even lower \citep{Claeyssens2026}. The field is rapidly evolving, as evidenced by the sheer number of papers that have appeared in the literature on this topic in the last year alone, and thus, this is merely a jumping off point to understand the formation of one of the most enigmatic populations that have long lurked in the halos and at the hearts of galaxies.

\section*{Acknowledgments}
K.E.W. gratefully acknowledges insightful conversations with Angela Adamo and Matteo Messa, and sincerely thanks Joel Pfeffer for sharing their data compilation and models. We thank the anonymous referee for providing insightful comments, which improved the clarity and quality of this manuscript.  This work is based in part on observations made with the NASA/ESA/CSA James Webb Space Telescope and the NASA/ESA Hubble Space Telescope obtained from the Space Telescope Science Institute, which is operated by the Association of Universities for Research in Astronomy, Inc., under NASA contract NAS 5–26555. The data were obtained from the Mikulski Archive for Space Telescopes at the Space Telescope Science Institute, which is operated by the Association of Universities for Research in Astronomy, Inc., under NASA contract NAS 5-03127 for JWST. These observations are associated with programs JWST-GO-2561, JWST-GO-4111, JWST-ERS-1324, JWST-DD-2756. HST-GO-11689, HST-GO-13386, HST-GO/DD-13495, HST-GO-13389, HST-GO-15117, and HST-GO/DD-17231. 
Financial support for programs JWST-GO-2561, JWST-GO-4111, and JWST-GO-6405 is gratefully acknowledged and is provided by NASA through grants from the Space Telescope Science Institute, which is operated by the Associations of Universities for Research in Astronomy, Incorporated, under NASA contract NAS 5-03127.  Some of the data products presented herein were retrieved from the Dawn JWST Archive (DJA). DJA is an initiative of the Cosmic Dawn Center, which is funded by the Danish National Research Foundation under grant No. 140.
A.Z. acknowledges support by Grant No. 2020750 from the United States-Israel Binational Science Foundation (BSF) and Grant No. 2109066 from the United States National Science Foundation (NSF); and by the Israel Science Foundation Grant No. 864/23.
J.R.W. acknowledges that support for this work was provided by The Brinson Foundation through a Brinson Prize Fellowship grant.

\facilities{JWST (NIRCam)}

\software{
\textsc{astropy} \citep{astropy2013,astropy2018,astropy2022},
\eazy{} \citep{Brammer2008}
\grizli{} \citep[\url{github.com/gbrammer/grizli}]{grizli},
\galfit{} \citep{Peng2002,Peng2010},
\pros{} \citep{prospector2021},
\textsc{fsps} \citep{fsps2009,fsps2010a,fsps2010b}
\textsc{python-fsps} \citep{pythonfsps2014}
\textsc{aperpy} \citep[\url{github.com/astrowhit/aperpy}]{Weaver2024}
\textsc{sextractor} \citep{Bertin1996},
\textsc{sep} \citep{Barbary2016},
\textsc{extinction} \citep{extinction},
\textsc{sfdmap} \citep[\url{github.com/kbarbary/sfdmap}]{Schlegel1998,Schlafly2011},
\textsc{pypher} \citep{Boucaud2016},
\textsc{photutils} \citep{photutils2022},
\textsc{astrodrizzle} \citep{Gonzaga2012},
\textsc{numpy} \citep{numpy2011},
\textsc{matplotlib} \citep{matplotlib2007}
\textsc{pysersic} \citep{Pasha2023}
}

\bibliographystyle{aasjournal}
\bibliography{references}
\end{document}